\documentclass{aa}  
\usepackage{adjustbox}
\usepackage{color}
\usepackage{xcolor}
\usepackage{graphicx}
\usepackage{amsmath}
\usepackage{amssymb}
\usepackage{breqn}
\usepackage{gensymb}
\usepackage{placeins}
\usepackage{natbib}
\usepackage{upgreek}
\usepackage{subfig}
\usepackage{multirow}
\usepackage{booktabs}
\usepackage{colortbl}
\usepackage{csvsimple}
\usepackage[colorlinks=true,
linkcolor=blue,
citecolor=magenta]{hyperref}
\usepackage{float}
\usepackage{txfonts}
\usepackage{comment}

\usepackage[switch]{lineno}

\begin{document} 
\title{Probing spacetime and accretion model for the Galactic Center: Comparison of Kerr and dilaton black hole shadows}
\titlerunning{Probing spacetime and accretion model for the Galactic Center}
\authorrunning{R\"oder et. al.}

\author{Jan R\"oder \inst{1,2}\fnmsep\thanks{jroeder.astro@gmail.com\\
${}^{\star}$ Member of the International Max Planck Research School (IMPRS) for Astronomy and Astrophysics at the Universities of Bonn and Cologne} ~, Alejandro Cruz-Osorio \inst{2}\fnmsep\thanks{osorio@itp.uni-frankfurt.de}, Christian M. Fromm \inst{3,2,1}, Yosuke Mizuno \inst{4,5,2}, Ziri Younsi \inst{6,2}, and Luciano Rezzolla \inst{2,7,8}
}

\institute{    
	      Max-Planck-Institut f\"ur Radioastronomie, Auf dem H\"ugel 69, D-53121 Bonn, Germany
	      \and 
          Institut f\"ur Theoretische Physik, Max-von-Laue-Stra{\ss}e 1, D-60438 Frankfurt am Main, Germany
          \and 
          Julius-Maximilians-Universität Würzburg, Emil-Fischer-Straße 31, 97074 Würzburg 
          \and 
          Tsung-Dao Lee Institute, Shanghai Jiao Tong University, Shanghai, 201210, China
          \and 
          School of Physics and Astronomy, Shanghai Jiao Tong University, Shanghai, 201240, China
          \and 
          Mullard Space Science Laboratory, University College London, Holmbury St. Mary, Dorking, Surrey, RH5 6NT, UK
          \and 
          Frankfurt Institute for Advanced Studies, Ruth-Moufang-Stra{\ss}e 1, 60438 Frankfurt am Main, Germany
          \and 
          School of Mathematics, Trinity College, Dublin 2, Ireland
          }

   \date{Received September 02, 2022; accepted January 20, 2023}

\abstract
{
 	In the 2017 observation campaign, the Event Horizon Telescope (EHT) for the first time gathered enough data to image the shadow of the super-massive black hole (SMBH) in M\,87. Most recently in 2022, the EHT has published the results for the SMBH at the Galactic Center, Sgr\,A$^\ast$.
 	In the vicinity of black holes, the influence of strong gravity, plasma physics, and emission processes govern the behavior of the system. Since observations such as those carried out by the EHT are not yet able to unambiguously constrain models for astrophysical and gravitational properties,  
 	it is imperative to explore the accretion models, particle distribution function, and description of the spacetime geometry. Our current understanding of these properties is often based on the assumption that the spacetime is well-described by by the Kerr solution to general relativity, combined with basic emission and accretion models.
 	We explore alternative models for each property performing general relativistic magnetohydrodynamic and radiative transfer simulations.
	}
{ 
	By choosing a Kerr solution to general relativity and a dilaton solution to Einstein-Maxwell-dilaton-axion gravity as exemplary black hole background spacetimes, we aim to investigate the influence of accretion and emission models on the ability to distinguish black holes in two theories of gravity.
}
{
	We carry out three-dimensional general relativistic magnetohydrodynamics (GRMHD) simulations of both black holes, matched at their innermost stable circular orbit, in two distinct accretion scenarios. Using general-relativistic radiative transfer (GRRT) calculations, we model the thermal synchrotron emission and in the next step apply a non-thermal electron distribution function, exploring representative parameters to compare with multiwavelength observations. We further consider Kerr and dilaton black holes matched at their unstable circular photon orbits, as well as their event horizons.
}
{
	From the comparison of GRMHD simulations, we found a wider jet opening angle and higher magnetisation in the Kerr spacetime. Generally, MAD models showed larger magnetic flux than SANE, as is expected. The GRRT image morphology shows differences between spacetimes due to the Doppler boosting in the Kerr spacetime. However, from pixel-by-pixel comparison we find that in a real-world observation an imaging approach may not be sufficient to distinguish the spacetimes using the current finite resolution of the EHT. From multiwavelength emission and spectral index analysis, we find that accretion model and spacetime have only a small impact on the spectra compared to the choice of emission model. Matching the black holes at the unstable photon orbit or the event horizon further decreases the observed differences.
}
{}
	%


\maketitle

\section{Introduction}\label{sec:intro}

In April 2019, the Event Horizon Telescope Collaboration (EHTC) published the first ever image of the supermassive black hole (SMBH) at the heart of M87 \citep{EHT_M87_PaperI,EHT_M87_PaperII,EHT_M87_PaperIII,EHT_M87_PaperIV,EHT_M87_PaperV,EHT_M87_PaperVI}. The 2017 observation campaign included numerous other targets, including the Galactic Center black hole Sgr\,A$^\ast$. The existence of an SMBH in the center of our own galaxy has been inferred from stellar dynamics observations over multiple decades by two independent groups \citep{Ghez:2008,Gillessen:2009}. In the wake of the publication of the first EHT radio image of Sgr\,A$^\ast$ \citep{EHT_SgrA_PaperI,EHT_SgrA_PaperII,EHT_SgrA_PaperIII,EHT_SgrA_PaperIV,EHT_SgrA_PaperV,EHT_SgrA_PaperVI}, open questions concern the type of compact object residing in the Galactic Center. At the same time, accretion physics onto compact objects and emission processes in the direct vicinity around them are not fully understood. In the past, numerous theoretical comparisons between background spacetimes and their effect on the photon ring size \citep{Wielgus2021,Cruz-Osorio2021a}, black hole charge \citep{Kocherlakota2021} and synthetic images have been carried out \citep{Lu2014,Lu2016,Olivares2020,Mizuno2018,Fromm2021a,Younsi2021,Ozel2021,Kocherlakota2022}.  Still, tests of general relativity in the strong-field regime pose an immense challenge, especially once effects such as plasma turbulence and limited telescope resolution come into play. 
This work aims to compare two fundamentally different spacetimes and various combinations of accretion and emission models. We build on the pioneering work of \cite{Mizuno2018}, who compared a moderately rotating Kerr to a non-rotating dilaton black hole in a full-3D GRMHD simulation. First, we expand on this work by including a Magnetically Arrested Disk (MAD) accretion model in addition to the Standard and Normal Evolution (SANE).
From the GRMHD simulations, we find that the accretion torus consists of hot low-density collision-less ion plasma. Since we are interested in the radiating electrons, a bridging function for the respective temperatures is required. Many works such as \cite{Mizuno2018} used a constant value to relate the two; however, electrons radiate efficiently and Coulomb-collision ion cooling is suppressed due to low densities and high temperatures. Additionally, various heating processes are at play \cite{Chael2018,Mizuno2021}. Due to the different impacts of the aforementioned processes, the electron and proton temperatures are believed to be non-linearly related. Since our simulations do not include electrons, we employ a parametrization of the electron temperature as an effective model in radiative post-processing \citep{Moscibrodzka2016}. Further, plasma processes such as magnetic reconnection and instabilities lead to acceleration of electrons in some regions around black holes. Therefore, we employ a non-thermal electron energy distribution function \citep[see, e.g.][]{Ball2018a,Davelaar2018,Davelaar2019,Cruz-Osorio2021b}. Following \cite{Mizuno2018}, we also consider matching the Kerr and dilaton black holes at their unstable circular photon orbits and at their event horizons. Additionally, in order to disentangle effects of black hole spin from the presence of additional fields in the background spacetime, we include a Schwarzschild simulation. 

Other works investigating alternative theories of gravity \citep[e.\,g.][]{EHT_SgrA_PaperVI,Kocherlakota2021,Kocherlakota2022,Younsi2021,Ozel2021,Vagnozzi2022} often make use of semi-analytical plasma physics and emission models but include a broad spectrum of theories in their study. In this work, we restricted us to a single theory, the dilaton black hole, but aimed to create a scenario more akin to reality. The dilaton parameter used in this study is well within the current constraints \citep[see Fig. 18 of][]{EHT_SgrA_PaperVI}. Instead of semi-analytic models, we use state-of-the-art GRMHD and GRRT simulations to model accretion and emission physics in this alternative theory of gravity.

In this study, we scale our simulations to Sgr~A$^\ast$ (RA 17h 45m 40s, Dec -29$^\circ$ 0' 28'', \citealt{Petrov2011}) as a representative system. We use a mass of $M_{\rm BH}=4.148\times10^{6}$\,M$_\odot$, at a distance of $D_{\rm BH}=8.175$\,kpc \citep{Gravity2019}. 
The paper is structured as follows: In Section \ref{sec:methods} we describe the setup of the GRMHD simulations and GRRT calculations. We present the results in the same order in Section \ref{sec:results} along with a spectral analysis, and discuss them in Section \ref{sec:summary}. We present our final conclusions in Section \ref{sec:conclusion}.

\section{Methods}\label{sec:methods}

\subsection{General-Relativistic Magneto-hydrodynamics (GRMHD)}\label{sec:GRMHD}

In this work, we investigate two exemplary black hole systems in full GRMHD. Following the setup in \cite{Mizuno2018}, we choose a Kerr black hole with dimensionless spin $a_\star=0.6$ and a dilaton black hole with dilaton parameter $\hat{b}=0.504$ in spherically symmetric polar coordinates. The value of $\hat{b}$ is consistent with constraints obtained in recent studies \citep{EHT_SgrA_PaperVI,Kocherlakota2021} and quantifies a deviation from GR through a contribution to the black hole mass caused by the presence of the dilaton.
The dilaton black hole is described by Einstein-Maxwell-dilaton-axion (EMDA) gravity, which in turn has its roots in the low-energy effective formulation of string theory \citep{Garcia1995}. In EMDA gravity, the scalar dilaton and axion vector fields couple to the Faraday tensor. 

In order to arrive at two comparable systems with similar plasma dynamics, the black holes are matched to have the same innermost stable circular orbit (ISCO). Likewise, event horizon or photon ring can be chosen to mach the spacetimes. To this end, the dilaton parameter was calculated from the Kerr black hole's spin by equating the respective expressions for the ISCO. The analytic expressions of the dilaton metric along with charactersitic radii and the resulting matchings are reported in appendix \ref{sec:matchings_app}. For more details on EMDA gravity, see e.\,g. \citep{WeiLiu2013,Flathmann2015,Banerjee2021a,Banerjee2021b}.

It is important to note that instead of simply highlighting differences between two objects, we show that even differences between to two fundamentally dissimilar objects are not appreciable at the moment. In the former case, it would be best to compare similar objects, e.\,g. a Schwarzschild to a non-rotating dilaton black hole. Since we however take the latter approach to this challenge, we consider the rotating Kerr and the non-rotating dilaton black hole. Additionally, given a fixed black hole mass, the Schwarzschild metric does not contain any degree(s) of freedom through which systems with a common ISCO, unstable photon orbit or event horizon radii could be explored.

This work does not aim to extensively investigate spacetime parameters or non-GR spacetimes in a general manner, but rather serves as a case study on the distinguishability of two different spacetimes. Moreover, we study plasma properties in the framework of accretion models, electron temperature, and electron distribution function by changing plasma parameters like density, emissivity and opacity.

In the EMDA metric, we set both the axion field and the spin of the dilaton black hole to zero \citep{Mizuno2018}. The metric then reduces to a Schwarzschild-like expression with the dilaton parameter $\hat{b}$ as the remaining degree of freedom, in a sense quantifying the deviation from general relativity. 

The fundamental GRMHD equations read \citep[e.\,g.][]{Rezzolla_book:2013,Porth2017}:
\begin{linenomath}\begin{equation} \label{eq:grmhd_eqs}
\nabla_\mu\left(\rho u^\mu\right)=0,\ \ \ \nabla_\mu T^{\mu\nu}=0,\ \ \ \nabla_\mu {}^{*}\!F^{\mu\nu}=0,
\end{equation}\end{linenomath}
and describe local conservation of mass, energy and momentum and Faraday's law. In the first equation, $\rho$ is the rest mass density and $u^\mu$ is the fluid four-velocity. The energy-momentum tensor $T^{\mu\nu}$ and the dual of the Faraday tensor $^{*}\!F^{\mu\nu}$ read:
\begin{linenomath}\begin{equation}
T^{\mu\nu}=\rho h_{\rm tot}u^\mu u^\nu +p_{\rm tot} g^{\mu\nu}-b^\mu b^\nu,\ \ \ ^{*}\!F^{\mu\nu}=b^\mu u^\nu - b^\nu u^\mu,
\end{equation}\end{linenomath}
where $p_{\rm tot}=p+b^2/2$ and $h_{\rm tot}=h+b^2/\rho$ are the total pressure and specific enthalpy, respectively.
The magnetic field strength in the fluid frame and magnetic field four-vector are denoted by $b^2=b^\mu b_\mu$ and $b^\mu$. 

In total, four GRMHD simulations of the Kerr and dilaton black holes with two distinct magnetic field configurations were carried out. The extents of the numerical grid and other parameters are summarized in Table \ref{tab:grmhd_params}. 

The setup is identical to \cite{Mizuno2018}: An initially stationary torus in hydrostatic equilibrium with a weak poloidal magnetic field is set up around the black hole. For both the Kerr and dilaton tori a constant specific angular momentum distribution is chosen, with $l_{\rm Kerr}=4.5$ and $l_{\rm dilaton}=4.567$. These values determine the inner edge of the torus to be $r_{\rm in}=10.3$\,M in both systems \citep{Font02b,Rezzolla_book:2013,Cruz2020}. Inside the outermost closed equipotential surface, thermodynamic quantities are computed with an ideal gas equation of state with an adiabatic index $\Gamma=4/3$. In order to avoid vacuum regions, floor values are applied whenever a cell satisfies $\rho\leq\rho_{\rm floor}=10^{-4} r^{-3/2}$ or $p \leq p_{\rm floor}=(10^{-6}/3) r^{-5/2}$ \citep{Mizuno2018}. Since the torus is stationary by construction, the magneto-rotational instability (MRI) is triggered by randomly perturbing the gas pressure by about 1\%. The MRI develops and subsequently drives the accretion process.

The vector potential of the poloidal magnetic field (with $q$ as defined below) has the general form 
\begin{linenomath}\begin{equation}
	A_\phi \propto \max\left(q-0.2,0\right),
\end{equation}\end{linenomath}
and is added on top of the constructed torus. In this work, both a weak and a strong magnetic field configuration are considered. The former will produce a Standard and Normal Evolution (SANE) scenario \citep{Narayan2012,Mizuno2018,Ripperda2020}, while the latter is likely to result in a Magnetically Arrested Disk (MAD) situation. For each case, 
\begin{linenomath}\begin{align} 
q &= \frac{\rho} {\rho_{\rm max}}, \   &\mathrm{for\ \ SANE}\\
q &= \frac{\rho} {\rho_{\rm max}}\left(\frac{r}{r_{\rm in}}\right)^3 \sin^3 \theta \ \exp \left(-\frac{r}{400}\right),  &\mathrm{for\ \ MAD}
\end{align}\end{linenomath}
see also \cite{Fishbone76,Font02b,Rezzolla_book:2013}.
In SANE, matter can continuously accrete onto the black hole since the magnetic field is weak and disordered \citep{Narayan2012}. In the strong field case, more magnetic flux can pile up near the black hole, blocking off accretion and thereby "arresting" the disk \citep{Narayan2012}.

The simulation domain spans $r\in\left(0.8\,r_{\rm eh}, 1,000\,{\rm M}\right)$, $\theta\in\left(0.01\pi, 0.99\pi\right)$ and $\phi\in\left(0, 2\pi\right)$, covered by the numerical grid with $(N_r,N_\theta,N_\phi)=\left(256,128,128\right)$ gridpoints. The grid is logarithmic in the radial direction to naturally grant a higher resolution near the black hole, and uniform in the azimuthal and polar directions. At the inner and outer radial boundaries, standard inflow and outflow boundary conditions are employed, respectively. Along the polar boundaries, a solid reflective wall sets the flux through it to zero \citep{Shiokawa2012,Mizuno2018}. In the azimuthal direction, boundary conditions are simply periodic. 

The GRMHD equations \eqref{eq:grmhd_eqs}, in conservative and 3+1 split form, are solved by the \texttt{Black Hole Accretion Code BHAC} \citep{Porth2017}. It is a multidimensional extension to the \texttt{MPI-AMRVAC} framework \citep{Porth2014,Keppens2012} capable of evolving the GRMHD equations in a finite volume representation, in arbitrary spacetimes and coordinates. For a comparison with other state-of-the-art GRMHD codes, see \cite{Porth2019_etal}. 

In our setup, grid cell-interface values of primitive variables are calculated using a piecewise-parabolic method, resulting in local Riemann problems handled by a total variation diminishing Lax-Friedrichs scheme. For the time advance, a predictor-corrector scheme is employed \citep{Porth2017}. The spherically symmetric dilaton metric is implemented in Rezzolla-Zhidenko (RZ) parametrized form \citep{Rezzolla2014,Konoplya2016a}. 

Even though \texttt{BHAC} is capable of adaptive mesh refinement (AMR), our setup does not make use of it, enabling us to handle conservation of the $\boldsymbol{\nabla\cdot B} =0$ constraint using flux-interpolated constrained transport (FCT) \citep{Olivares2019}. We further employ modified Kerr-Schild coordinates, along with their parametrized form for the dilaton system \citep{Porth2017,Mizuno2018}. \\

\begin{table*}[h!]
	\vspace{-0pt}
	\centering
	\def\arraystretch{1.2}
	\caption{GRMHD parameters for SANE and MAD simulations, adapted from \cite{Mizuno2018}, $r_{\rm eh}$ is the event horizon radius.}
	
	\begin{tabular}{llll} 
		\hline \hline 
		\multicolumn{4}{c}{Plasma}\\
		adiab. index $\Gamma$ & density floor $\rho_{\rm fl}$ & pressure floor $p_{\rm gas,\ fl}$  & accretion model\\
		$4/3$ & $10^{-4} r^{-3/2}$ &$ (10^{-6}/3) r^{-5/2}$ & MAD, SANE\\
		\hline 	
		
		\multicolumn{4}{c}{Spacetime$^\ast$}\\
		$a$ & $\hat{b}$ & $l_{\rm torus,\ Kerr}$ & $l_{\rm torus,\ dilaton}$ \\
		$0.6$ & $0.504$ & $4.5$ & $4.567$ \\
		\hline 		
		
		\multicolumn{4}{c}{Grid extent}\\
		radial, $r$ & azimuthal, $\theta$ & polar, $\phi$ & cells, $(N_r,N_\theta,N_\phi)$\\
		$ \left(0.8\,r_{\rm eh}, 1,000\,{\rm M}\right)$ &$ \left(0.01\pi, 0.99\pi\right)$ &$\left(0, 2\pi\right) $&$\left(256,128,128\right)$ \\
		\hline 
		
		\multicolumn{4}{l}{$^\ast$ Kerr dimensionless spin parameter, dilaton parameter and specific angular momenta} \\
		\label{tab:grmhd_params}
	\end{tabular} 

\end{table*}

\subsection{Electron temperature and distribution function}\label{sec:ET}
In hot low-density plasmas, temperatures of ions and electrons are generally not equal, 
resulting in a two-temperature state \citep[][and references therein]{Yuan2014}. Since our GRMHD simulation only evolves the dynamically important protons, they have to be linked to the radiating electrons. \cite{Mizuno2018} used a constant proton-to-electron temperature ratio $T_{\rm p}/T_{\rm e}=3$; in this study, $T_{\rm p}/T_{\rm e}$ is set by a parametrization depending on the plasma parameter $\beta$ and an additional free parameter $R_{\rm high}$ \citep{Moscibrodzka2016}, where $\beta\equiv p_{\rm\,gas}/p_{\rm\,mag}$ is the ratio of gas-- to magnetic pressure. The $T_{\rm p}/T_{\rm e}$ parametrization is defined as:
\begin{linenomath}\begin{equation} \label{eq:tratio_beta}
	\frac{T_{\rm p}}{T_{\rm e}}=\frac{R_{\rm low}+R_{\rm high}\,\beta^2}{1+\beta^2}\,.
\end{equation}\end{linenomath}
For alternative electron temperature prescriptions, see \cite{Anantua2020}. The free parameters $R_{\rm high}$ and $R_{\rm low}$ control the temperature ratio in the disk ($\beta\gg1$) and in the jet ($\beta\ll1$). Throughout this work, the simplified version of the parametrization characterized by $R_{\rm low}=1$ is employed \citep{EHT_M87_PaperV}.
The electron temperature in cgs units is then calculated as
\begin{linenomath}\begin{equation}\label{eq:dim_less_T_e}
	T_{\rm e}=\frac{m_{\rm e}c^2}{k_{\rm B}} \Theta_{\rm e} = \frac{m_{\rm e}c^2}{k_{\rm B}} \Theta_{\rm p} \frac{m_{\rm p}}{m_{\rm e}} \left(\frac{T_{\rm p}}{T_{\rm e}}\right)^{-1}.
\end{equation}\end{linenomath}
$\Theta_{\rm e} \equiv k_{\rm B} T_{\rm e}/m_{\rm e} c^{2}$ is the dimensionless electron temperature; its proton (ion) equivalent $\Theta_{\rm p}$ is known from the GRMHD simulation. The R-$\beta$ parametrization has been shown to well model the presence of turbulent and magnetic reconnection heating of electrons \citep{Mizuno2021,Chael2018}. 
 
In addition to the thermal electron distribution function, a non-thermal kappa model is adopted \citep[e.\,g.][or, for recent application, \citealt{Davelaar2018,Davelaar2019}]{Vasyliunas1968,Tsallis1988,Tsallis1998,LivadotisMcComas2009}. All formulations of electron energy distribution functions, absorptivities and emissivities are taken from \cite{Pandya2016}. 
The thermal distribution function reads \citep{Leung2011}: 
\begin{linenomath}\begin{equation}\label{eq:rel_thermal}
	\frac{dn_{\rm e}}{d\gamma_{\rm e}\,d\cos\xi\,d\phi} = \frac{n_{\rm e}}{4 \pi \Theta_{\rm e}} \frac{\gamma_{\rm e} \left(\gamma_{\rm e}^2 - 1\right)^{1/2}}{K_2\left(1/\Theta_{\rm e}\right)} \exp \left(- \frac{\gamma_{\rm e}}{\Theta_{\rm e}}\right),
\end{equation}\end{linenomath}
with electron number density $n_{\rm e}$, gyrophase $\phi$, Lorentz factor $\gamma_{\rm e}$, pitch angle $\xi$ and modified Bessel function of the second kind $K_2$. The kappa distribution function can be written as \citep{Xiao2006}:
\begin{linenomath}\begin{equation} \label{eq:rel_kappa}
	\frac{dn_{\rm e}}{d\gamma_{\rm e}\,d\cos{\xi}\,d\phi} = \frac{N}{4 \pi} \gamma_{\rm e} \left(\gamma_{\rm e}^2 - 1\right)^{1/2} \left(1 + \frac{\gamma_{\rm e}-1}{\kappa w}\right)^{-(\kappa+1)},
\end{equation}\end{linenomath}
with normalization factor $N$ \citep{Pandya2016}. The kappa index is related to the high-energy power law slope $s$ as $\kappa=s+1$. The width of the distribution is explained below. 

In this work, $\kappa$ is not set to a constant value but is calculated from fluid variables based on particle-in-cell simulations of magnetic reconnection in current sheets \citep{Ball2018a}:
\begin{linenomath}\begin{equation}
	\kappa= 2.8+0.7\,\sigma^{0.5}+3.7\,\sigma^{-0.19}\,\tanh \left(23.4\,\sigma^{0.26}\,\beta\right),
\end{equation}\end{linenomath}
In the above equation, $\sigma=b^2/\rho$ is the magnetization. For different $\beta$ and $\sigma$, analytical fitting functions are obtained for $10^{-4}<\beta<1.5$ and $0.1<\sigma<7.2$. These values are believed to be consistent with typical values found in the outer jet wall.
The width of the distribution $w$ can be written to contain a thermal and a magnetic energy term \citep{Davelaar2019}:
\begin{linenomath}\begin{equation}\label{eq:w}
w=\frac{\kappa-3}{\kappa}\left(\Theta_{\rm e}+\frac{\varepsilon}{2}\left[1+\tanh\left(r-r_{\rm inj}\right)\right]\frac{m_{\rm p}}{m_{\rm e}} \frac{\sigma}{6}\right),
\end{equation}\end{linenomath}
where $\varepsilon$ sets the fraction of the magnetic energy contribution to the electron temperature. We set $\varepsilon$ to $0$ and $0.015$.\\

\subsection{General Relativistic Radiative Transfer (GRRT)}\label{sec:GRRT}

In order to model millimeter and sub-millimeter synchrotron emission, GRRT calculations are carried out on the GRMHD simulations. First, null geodesics (light rays) are integrated directly between the black hole system and a far-away observer. Then, the differential equations for intensity and optical depth are integrated along each ray \citep{Younsi2012}. They read:

\begin{equation} 
\frac{d\tau_\nu}{d\lambda} = \xi^{-1}\,\alpha_{0,\nu} \vphantom{\frac12}\,,\quad
\frac{d\mathcal{I}}{d\lambda} = \xi^{-1}\left(\frac{j_{0,\nu}}{\nu^3}\right)e^{-\tau_\nu}\vphantom{\frac12}\label{eq:int_ode_I},
\end{equation}
with affine parameter $\lambda$, optical depth $\tau_\nu$, Lorentz invariant intensity $\mathcal{I}$, frequency $\nu$, absorptivity $\alpha_{0,\nu}$ and emissivity $j_{0,\nu}$. For the latter two, subscript 0 denotes measurement in the rest frame.

In this work, we make use of the code \texttt{Black Hole Observations in Stationary Spacetimes BHOSS} \citep{Younsi2020}. The geodesics are handled by a Runge-Kutta-Fehlberg integrator, solving the equations to fourth order and adjusting the step size using a fifth-order error estimate \citep{Fehlberg1969}. The intensity equations are integrated in an Eulerian scheme along each previously obtained light trajectory. In \texttt{BHOSS}, a far-away observer is initialized in the form of an image plane perpendicular to the line of sight (towards the black hole system). The full camera setup is reported in \cite{Younsi2016}. All generated images are averages over 101 snapshots taken between 11,000\,M and 12,000\,M of the GRMHD simulation. This time span corresponds to about six hours for Sgr A$^\ast$. For each emission model, we iterate the mass accretion rate $\dot{M}$ until an average flux of 2.5\,Jy at 230\,GHz is obtained \citep{Bower2019}. 
Recently, it has been shown in ultra-high spatial resolution GRMHD simulations that magnetic reconnection takes place in both jet and disk \citep[e.\,g.][]{Ripperda2020}. Nonetheless, we apply non-thermal emission only in a narrow region within the jet wall, consistent with existing literature \citep{Davelaar2018,Davelaar2019}.  
We neglect any emission from regions where $\sigma\geq\sigma_{\rm cut}=1$. Additionally, we employ a constraint on the Bernoulli parameter $Be=-hu_t\geq1.02$. Where $Be$ exceeds unity, the gas is unbounded, feeding jet and wind outflow \citep{Moscibrodzka2013}. Image extent and resolution are summarized in Table \ref{tab:bhoss_params}, along with mass of, and distance to the black hole. 

For moderate spin, the most recent EHT results favor models with lower inclinations and higher values of $R_{\rm high}$ for Sgr\,A$^\ast$ \citep{EHT_SgrA_PaperV}. The inclination used here was adapted from \cite{Mizuno2018} to maintain comparability of results. As a follow-up to Mizuno et al., this study was started long before these results were on the horizon. We investigated $R_{\rm high}=80$ and 160 in \cite{Roder2022}. However, at the inclination and field of view chosen, the SANE image morphology does not change for $R_{\rm high}\geq40$, while the MAD images stop changing already for $R_{\rm high}\geq10$.

\begin{table}[h]
	\vspace{-0pt}
	\centering
	\def\arraystretch{1.2}
	\caption{GRRT parameters. }
	\begin{tabular}{llll} 
		\hline \hline 
		
		\multicolumn{4}{c}{Images}\\
		pixels & FOV ($\upmu$as)& inclination (deg) &  $S_{\rm 230\,GHz}$ (Jy)\\
		1024 & 300 &60 &2.5\\
		\hline 		
		
		\multicolumn{4}{c}{Emission model}\\
		$R_{\rm low}$ & $R_{\rm high}$& eDF & $\varepsilon$ \\
		1& 1, 10, 20, 40 & thermal, non-thermal & 0.0, 0.015\\
		\hline 		
		
		\label{tab:bhoss_params}
	\end{tabular} 
\end{table} 

\begin{table} [h!]
	\centering 
	\def\arraystretch{1.2}
	\setlength{\tabcolsep}{2.5pt}
	\caption{Mean values and standard deviations for $\dot{M}$, $\Phi_{\rm BH}$, and $\psi=\Phi_{\rm BH}/\sqrt{\dot{M}}$, computed between 11\,000\,M to 12\,000\,M. Values for the Schwarzschild spacetime are taken from \cite{Fromm2021b}.}
	\begin{tabular}{ l  l  lll }
		\hline\hline
		Metric          & Model  & $\langle\dot{M}\rangle$ & $\langle\Phi_{\rm BH}\rangle$ & $\langle\psi\rangle$ \\
		\hline
		Kerr 	& SANE      &  $6.5\pm0.8$  &  $5.09\pm0.13$   &  $2.00\pm0.11$ \\
		Dilaton & SANE  	&  $5.2\pm0.7$  &  $2.42\pm0.07$   &  $1.06\pm0.07$ \\
		Schwarzschild & SANE & $0.36\pm0.03$  &  $0.60\pm0.02$ &  $0.99\pm0.05$ \\
		Kerr 	& MAD       &  $2.2\pm0.5$  &  $12.79\pm0.09$  &  $8.6\pm0.7$   \\
		Dilaton & MAD   	&  $2.5\pm0.4$  &  $12.15\pm0.13$  &  $7.8\pm0.7$   \\
		Schwarzschild & MAD & $5.0\pm0.7$  &  $32\pm1$ &  $14\pm1$ \\
		\hline
	\end{tabular}
	\label{tab:grmhd_mean_stdev}
\end{table}

\section{Results}\label{sec:results}

\subsection{GRMHD simulations}\label{sec:grmhd_results}

The four Kerr/dilaton and SANE/MAD model configurations are evolved until 15,000\,M. Since the Kerr and dilaton black holes were matched to have the same ISCO, the overall dynamical behavior is quite similar. Past 10,000\,M (SANE) or 11,000\,M (MAD), the systems begins to saturate and finally enters a quasi-steady state. The analysis of magnetization, $\sigma$, plasma $\beta$, Lorentz factor, $\gamma$ and electron temperature, $\Theta_{\rm e}$, as well as all GRRT calculations are therefore carried out on the interval between 11,000\,M and 12,000\,M, equivalent to an observation time of around six hours for Sgr\,A$^\ast$.
Time averages and corresponding standard deviations for $\dot{M}$, $\Phi_{\rm BH}$ and $\psi$ over this interval are listed in Table \ref{tab:grmhd_mean_stdev}. 

Within the full evolution time of 15,000\,M, the MAD simulations are approaching the characteristic value $\psi\approx10$ for a MAD state \citep{Tchekhovskoy2011}. This is however, consistent with $\psi_{\rm max}\approx 15$ for $a=0.9735$ \citep{Porth2019_etal}, since the Kerr black hole in this work is only moderately rotating. 

\begin{figure*}
	\centering
	\subfloat{\includegraphics[width=0.9\textwidth]{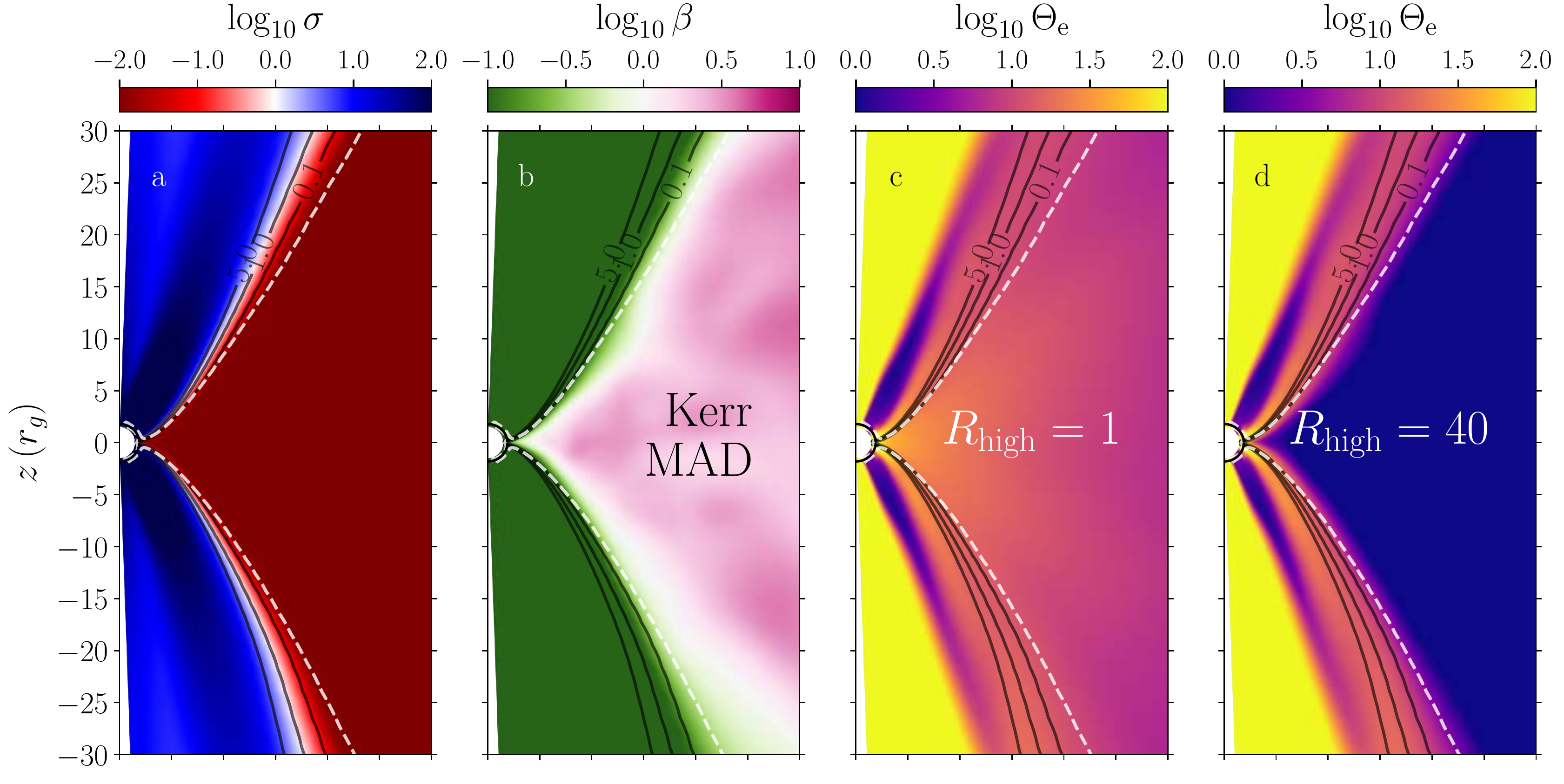}}  \\\vspace{-13pt}
	\subfloat{\hspace*{-0.055cm}\includegraphics[width=0.9\textwidth]{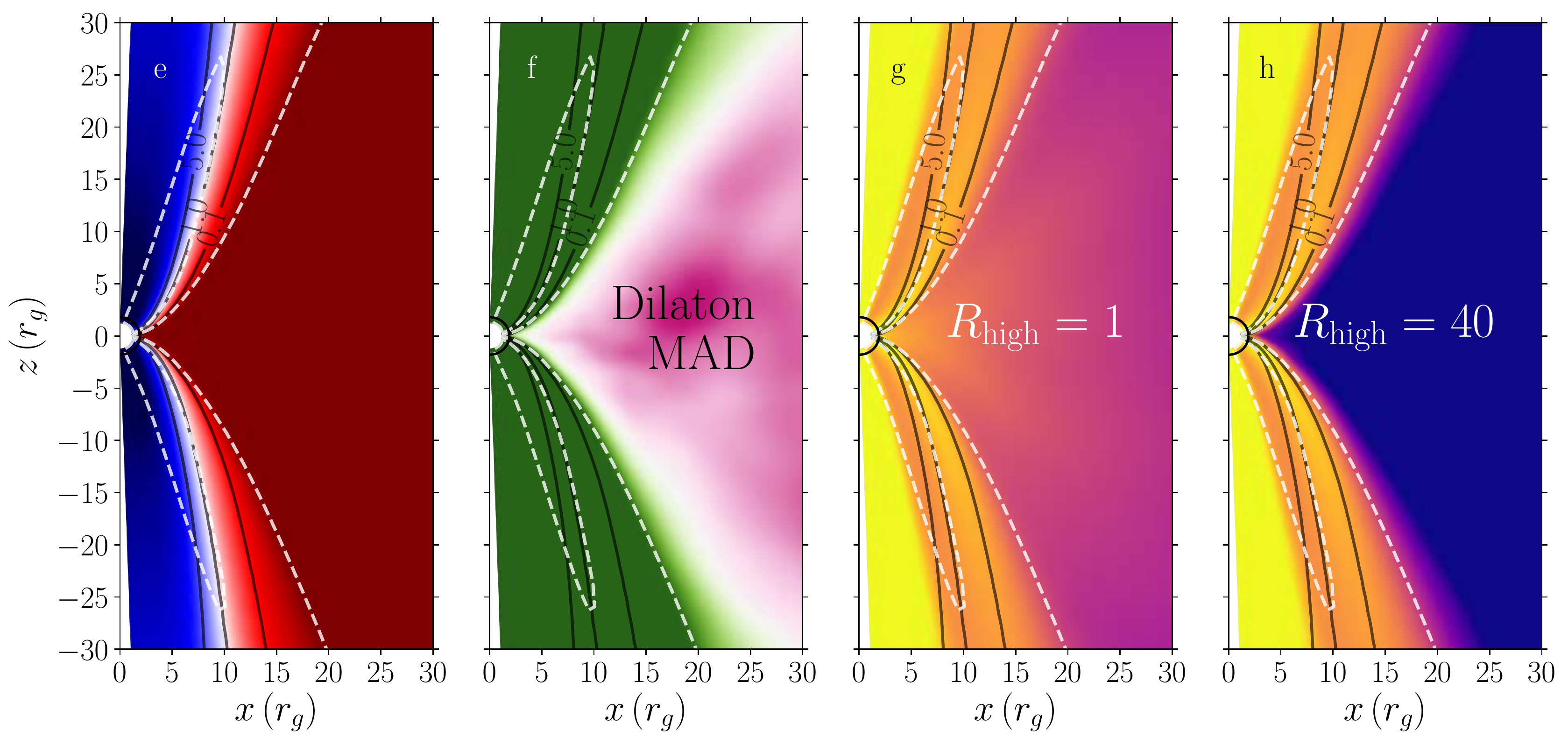}} \\
	\caption{Magnetization $\sigma$, plasma $\beta$ and electron temperature $\Theta_{\rm e}$ at $R_{\rm high}=1$ and 40 for MAD simulations in Kerr and dilaton spacetimes.  White dashed line: Bernoulli parameter $Be=1.02$. Annotated solid contour lines: levels of $\sigma$. The azimuthally averaged GRMHD data is shown time averaged over 1000\,M.} \label{fig:SBG2Te_MAD}
\end{figure*}

\begin{figure*}
	\centering
	\subfloat{\includegraphics[width=0.9\textwidth]{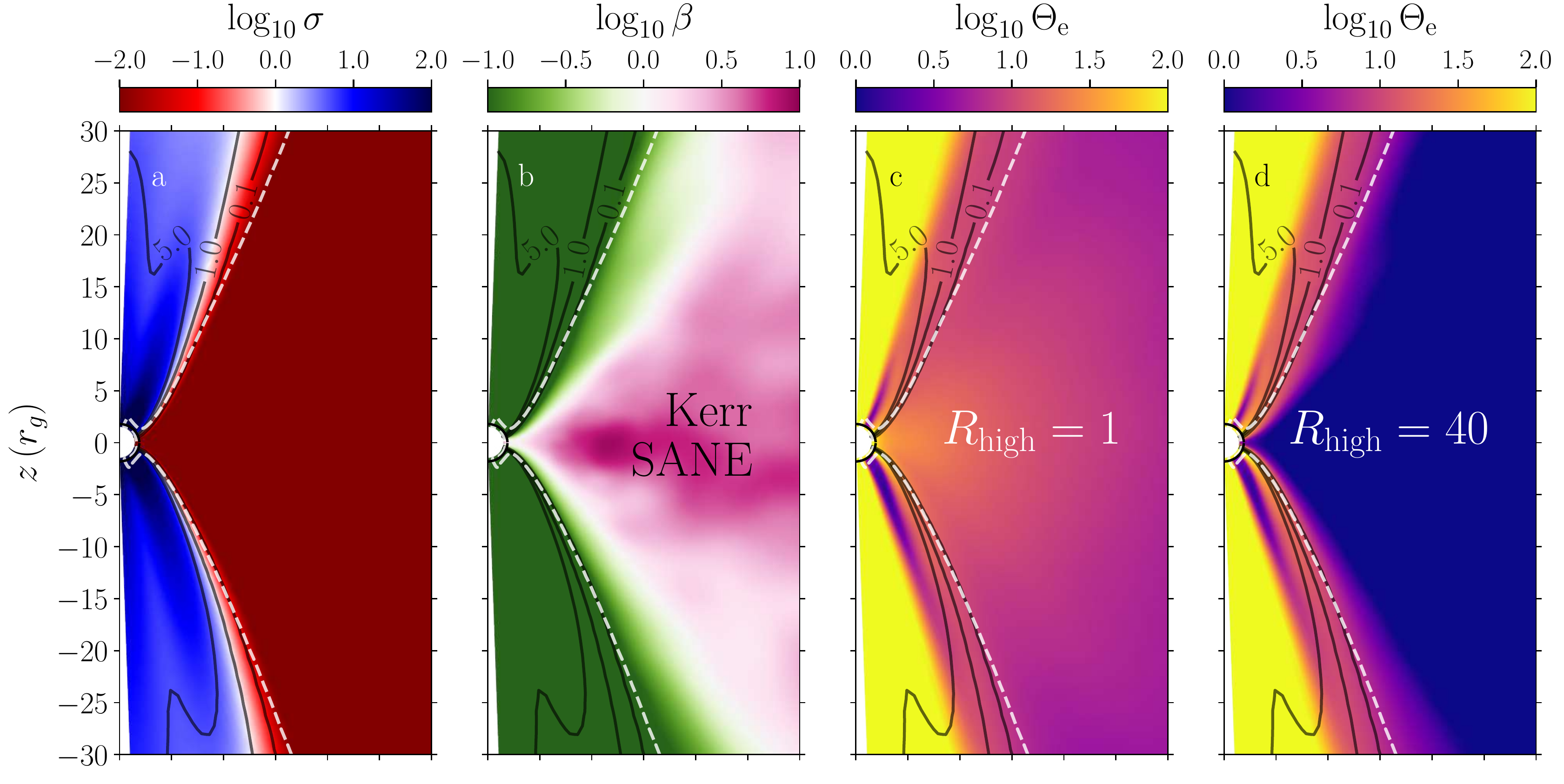}}  \\\vspace{-13pt}
	\subfloat{\hspace*{-0.055cm}\includegraphics[width=0.9\textwidth]{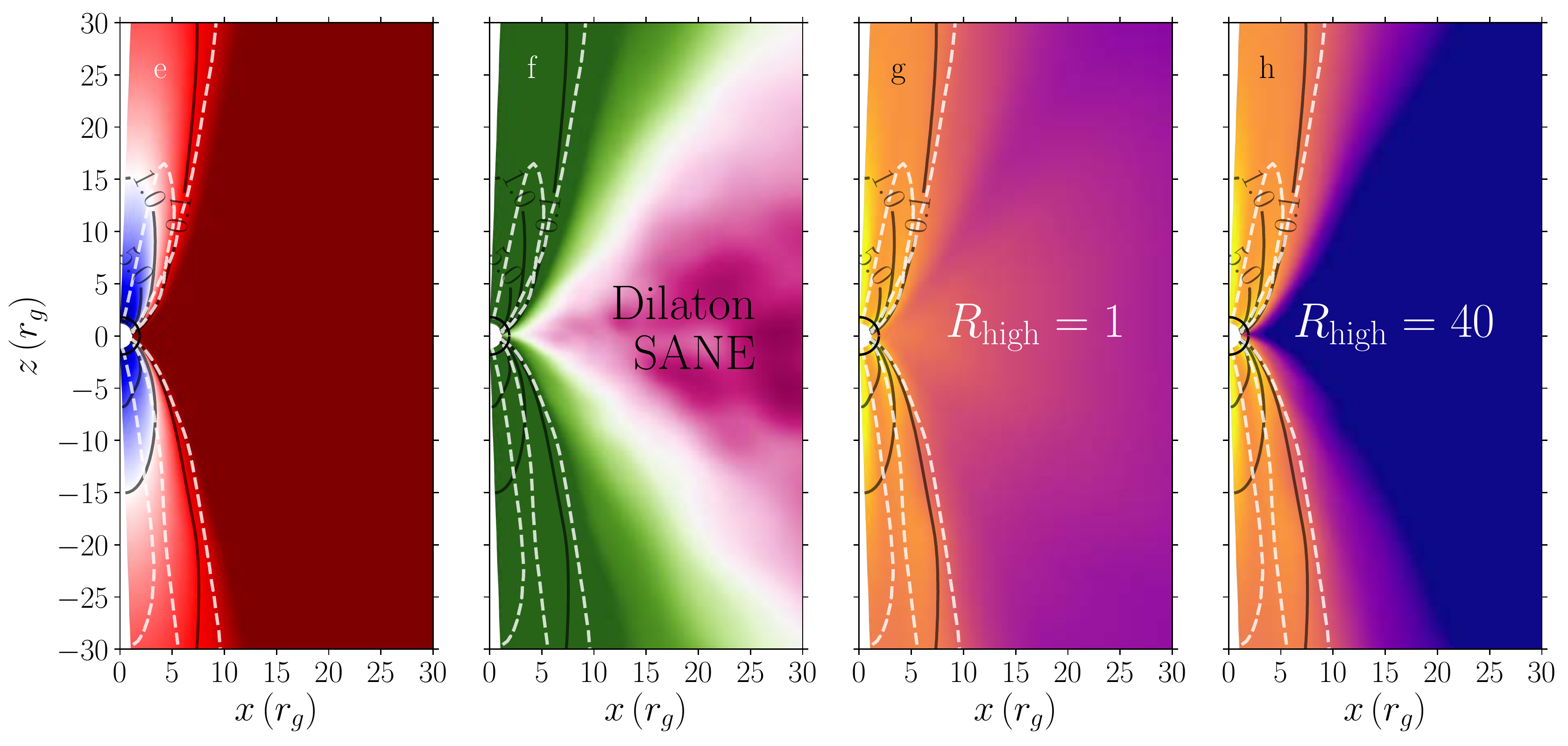}} \\
	\caption{Same as Fig. \ref{fig:SBG2Te_MAD}, but for the SANE simulation.} \label{fig:SBG2Te_SANE}
\end{figure*}

Figures \ref{fig:SBG2Te_MAD} and \ref{fig:SBG2Te_SANE} shows time and azimuthally averaged magnetization $\sigma$, plasma $\beta$ and electron temperature $\Theta_{\rm e}$ for two values of $R_{\rm high}$. The panels expand to 30\,$r_g$, corresponding to 150\,$\upmu$as in the GRRT images. We define here the jet spine to be bounded by $\sigma=1$ and the jet sheath as the region where $0.1<\sigma<1 \land Be<1.02$, where a Bernoulli parameter $Be>1$ describes unbounded gas feeding jet and wind outflow \citep{Moscibrodzka2013}.

In SANE, the torus in both spacetimes is weakly magnetized; this is a generic feature also present in the MAD simulation (panels a and e). The jet spine, however, is much more magnetized in Kerr than the corresponding regions in the dilaton system. For Kerr, in both SANE and MAD simulations, in agreement with Blandford-Znajek mechanism,  more magnetic flux accumulates near the horizon and the black hole's rotation causes an almost evacuated but highly magnetized separation region between sheath and spine, especially apparent in the Lorenz factor (see below). 

The Kerr system shows a much wider jet opening angle compared to the dilaton case. This can be seen at $|\,z\,|\,=30\,r_g$, where the outer edge of the sheath traced by $\sigma=0.1$ in the Kerr system extends out roughly twice as far in the $x$ direction (up to $15\,r_g$ for SANE and up to $21\,r_g$ for MAD) as in the dilaton system. The $Be=1.02$ contour line shows the same qualitative behavior. 
In SANE model, the highest magnetized region, where $\sigma\geq5$, is confined to the innermost  $\sim$$5\,r_g$ for the dilaton black hole, whereas in in the Kerr case, it stretches out five times as far. In the dilaton system even the $\sigma=1$ line wraps around the central region at $15\,r_g$ from the black hole (panel e), while in the Kerr system it extends into a similar direction as $\sigma=0.1$, following along the jet wall.

For both spacetimes (and both SANE and MAD accretion models), the $Be\geq1.02$ region shows a uniform distribution of low plasma $\beta$ (panels b, and f). In the dilaton torus (panel f), larger parts of the torus show higher values of $\beta$ near the mid-plane.  Through Eq. \eqref{eq:tratio_beta}, this difference in the distribution of $\beta$ in the torus plays an important role for the source morphology in the GRRT images (see below and \ref{sec:GRRT_images}), where the proton temperature is expected to be greater than electron temperature.

The Lorentz factor is generically low in the torus for both spacetimes. The aforementioned low-density separation region in the Kerr system is characterized by significantly higher Lorentz factors up to $\gamma\sim10$; this region is entirely absent in the dilaton system.
The MAD simulation further enhances the above-mentioned differences between Kerr and dilaton spacetimes. Both Kerr and dilaton jet opening angles are wider (e.g. $\sigma=0.1$ contour line), and now the dilaton black hole also shows a highly magnetized jet spine. However, in the now larger $Be\geq1.02$ region, plasma $\beta$ remains low. For both spacetimes, the distribution in the torus on the other hand shows much lower values compared to the SANE simulation. 

The two last columns of Figs. \ref{fig:SBG2Te_MAD} and \ref{fig:SBG2Te_SANE} show the dimensionless electron temperature, $\Theta_{\rm e}$, for $R_{\rm high}=1$ and 40 for SANE and MAD simulations. The evacuated separation region in the Kerr system appears as a particularly low electron temperature zone ($T_{\rm e}\sim T_{\rm p}$), extending $\sim$$10\,r_g$ outwards from the Kerr black hole in the SANE case. At low $R_{\rm high}$, the disk is filled with hot electrons ($\Theta_{\rm e}\sim10$). When $R_{\rm high}$ is increased, the electron temperature in the disk is decreased (compare e.g. panels c and d). The increase in $R_{\rm high}$ does not impact the temperature beyond $\sigma=0.1$ in the polar direction since we have fixed $R_{\rm low}=1$, for both SANE and MAD simulations of either spacetime. 

While the MAD simulation enhances the low-temperature appearance of the separation region in Kerr, the dilaton system also begins to show signs of such a region (panels c and g). At $R_{\rm high}=40$, the transition between low and high-temperature regions is sharper compared to the SANE simulation. In MAD, the Kerr jet sheath is moderately hotter, while the dilaton sheath is significantly hotter ($\Theta_{\rm e}\sim30$) than it is in SANE ($\Theta_{\rm e}\sim10$).

\subsection{GRRT images}\label{sec:GRRT_images}

\begin{figure*}
	\centering
	\includegraphics[width=0.83\textwidth]{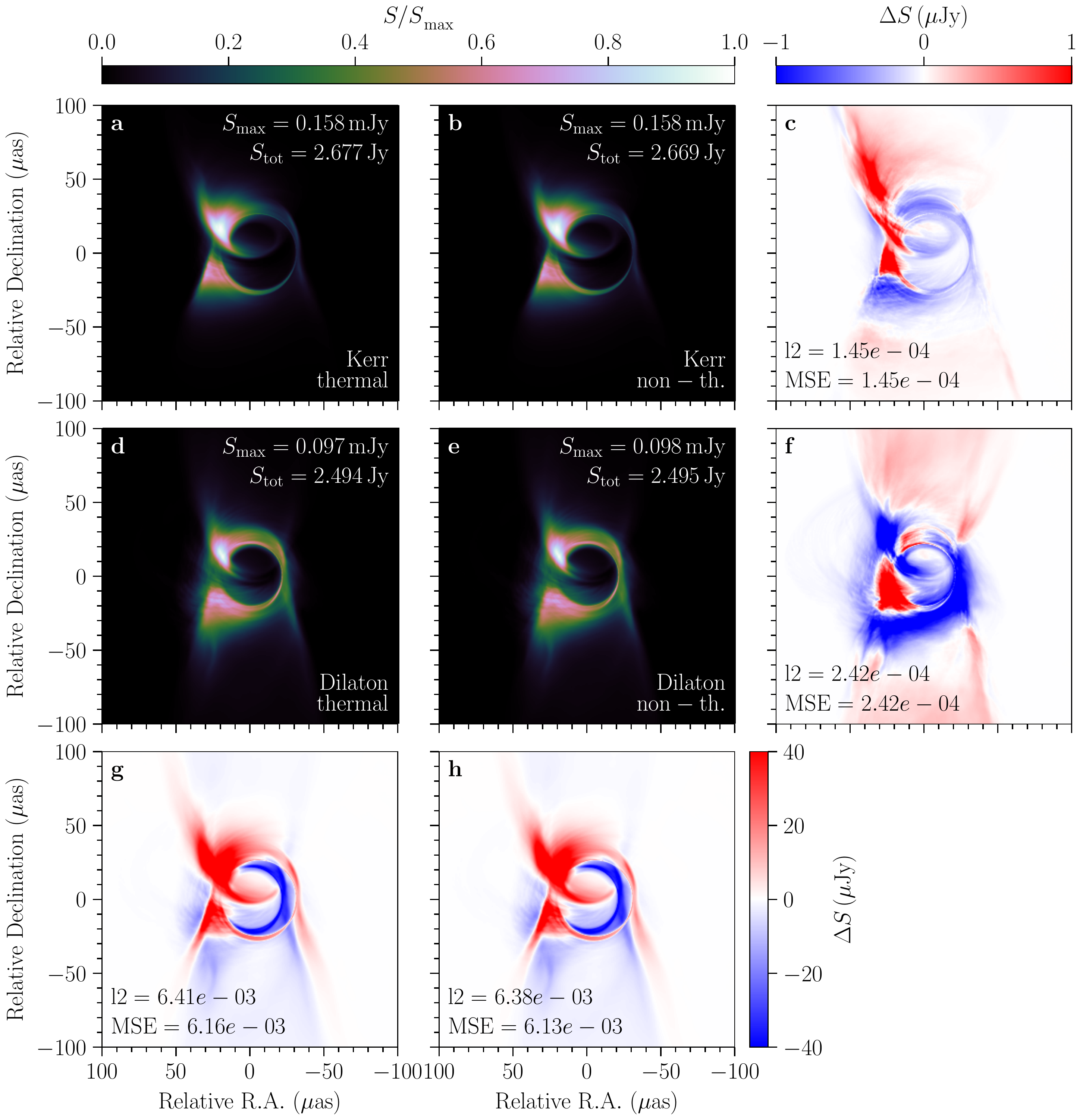}
	\caption{Kerr and dilaton GRRT images at $R_{\rm high}=40$ in the SANE simulation. The images are averages of 100 snapshots over 1\,000\,M simulation time ($\sim 6$\,hrs for Sgr\,A$^*$). This model configuration shows the largest difference between different electron distribution (eDF) functions (panel f) in the Dilaton spacetime in the given parameter space. The bottom row shows pixel-by-pixel differences between the two spacetimes at a given eDF.} \label{fig:edf_comp_SANE}
\end{figure*}

\begin{figure*}
	\centering
	\includegraphics[width=0.83\textwidth]{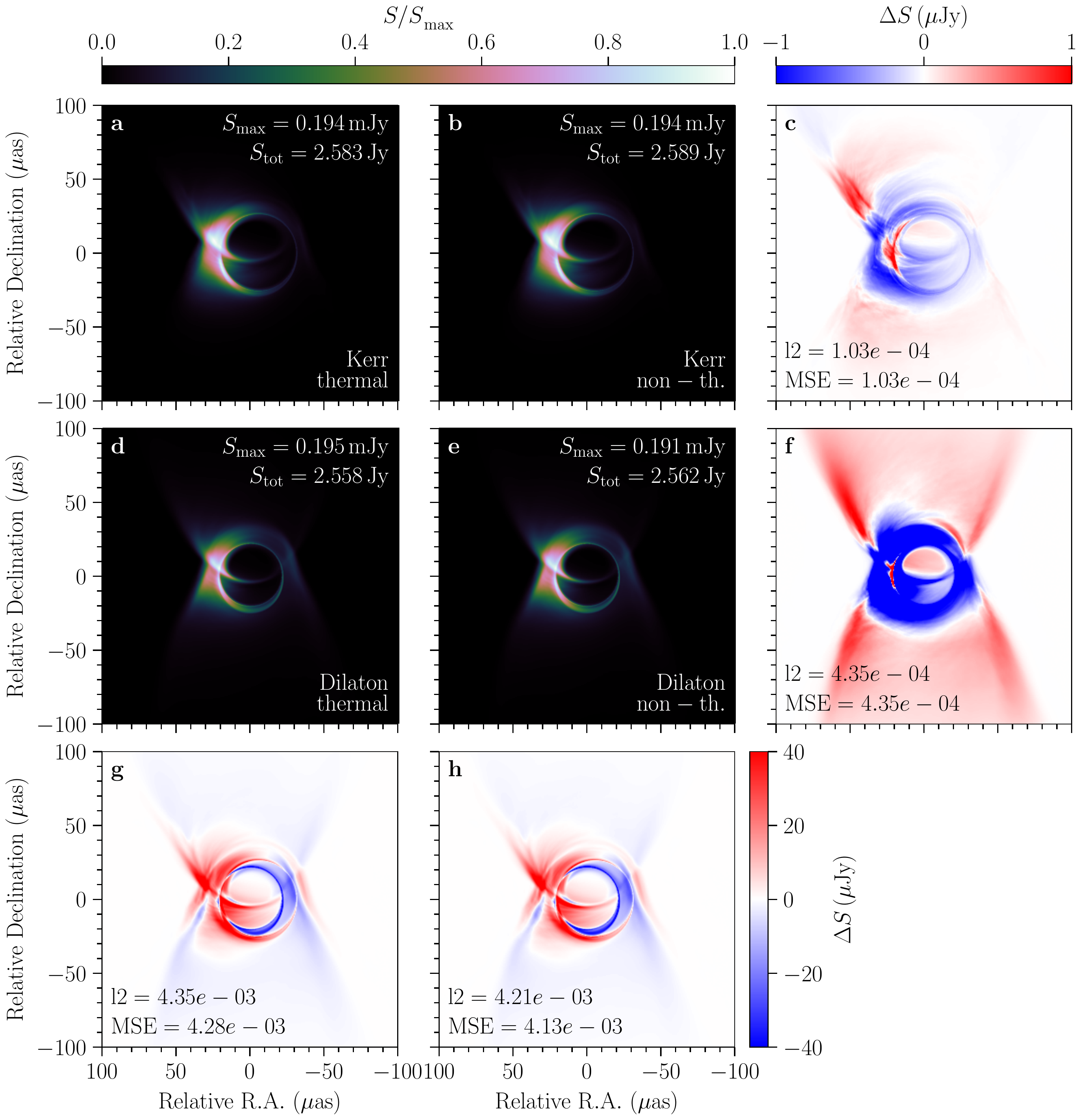}
	\caption{Same as Fig. \ref{fig:edf_comp_SANE}, but for the MAD case.} \label{fig:edf_comp_MAD}
\end{figure*}

Figures \ref{fig:edf_comp_SANE} and \ref{fig:edf_comp_MAD} show
time-averaged GRRT images for Kerr and dilaton black holes in SANE and
MAD simulations at 230\,GHz, with differences between electron
distribution functions in the rightmost column and differences between
the spacetimes at a given emission model in the bottom row. There is no 
visual difference between in source morphology whether non-thermal emission
is included or not for two reasons: one, the kappa model is applied only 
in a narrow region in the jet sheath, and two, we fix the flux at 230\,GHz and the kappa 
distribution shows its effects only at much higher energies (see Fig. \ref{fig:SED_SgrA}).
In the right column of Figs. \ref{fig:edf_comp_SANE} and \ref{fig:edf_comp_MAD}, the pixel-by-pixel 
differences between two images with different distribution functions are shown 
(the non-thermal image is subtracted from the thermal one). Intuitively, one may assume that the 
jet should be brighter in the non-thermal images and the torus should be dimmer; yet, the opposite is
the case. This can be explained by the shapes of the electron distribution functions: moving from a 
thermal to a non-thermal distribution, more electrons gain energy, shifting the maximum of the distribution 
and leaving the energy level we observe in the images with a lower number of electrons.

For any combination of accretion model with Kerr or dilaton spacetimes, 
the absolute difference between two corresponding pixels in
different electron distribution functions (eDFs) at 230\,GHz does not exceed
$5.5\upmu$Jy. Comparing the Kerr and dilaton spacetimes 
to a Schwarzschild one yields only marginally larger differences. In the Kerr spacetime, higher total flux is produced by non-thermal particles in the jet accelerated
 by the Blandford-Znajek mechanism. The total flux in the dilaton black system is 
 lower than for corresponding Schwarzschild simulations. For details, see Appendix \ref{sec:schwarzschild}.

\subsubsection{SANE simulation}

For $R_{\rm high}<10$, the Kerr and dilaton black holes show a very similar source morphology (panels a and d on the left side of Fig. \ref{fig:GRRT_ISCO}). The structure is highly asymmetric, with nearly all the flux concentrated in the left half of the image. The region of peak emission (down to 80\% of the peak flux, almost white) traces the left edge of the shadow at around zero relative declination in both spacetimes. It extends a few tens of micro-arcseconds outwards and is enclosed by the 60\% region (orange-red), which is still confined to the approaching side for $R_{\rm high}=1$. The same holds for the 40\% emission regime (green), wrapping half-way around the shadow and stretching out in a thin veil across it. At $R_{\rm high}=1$, the total fluxes in Kerr and dilaton images are identical, regardless of the chosen electron distribution function. 

When $R_{\rm high}$ is increased, the receding side in the dilaton spacetime begins to become more prominent, while in the Kerr system it stays faint. This is due to the Kerr black hole's ergosphere, where photons and matter are frame-dragged along with the spacetime in the direction of he black hole's rotation. This effect adds to the Doppler boosting (and thereby to the source asymmetry), brightening the approaching side and darkening the receding side. The dilaton black hole is non-rotating and therefore does not have an ergosphere, and the asymmetry is caused purely by Doppler boosting. 

Up until $R_{\rm high}=10$, the 60\% flux region in the dilaton torus stretches out much farther over the torus compared to the Kerr case, up to $\sim30\,\upmu$as from the left edge of the shadow (panels b and e in Fig. \ref{fig:GRRT_ISCO}). Nevertheless, the Kerr images are generically brighter than the dilaton images for $R_{\rm high}>1$ despite the fact that in the dilaton system more emission from the jet contributes to the total flux. This can be seen from the two rightmost columns in Fig. \ref{fig:SBG2Te_SANE}, where the $\sigma_{\rm cut}=1$ contour line traces a significantly larger region in the Kerr system from where we exclude all emission.

At $R_{\rm high}=20$, filamentary structures begin to stretch out from the torus in to the north and south directions. This gives the torus in both spacetimes a fuzzy appearance that develops into a more clearly defined jet onset upon further increase in $R_{\rm high}$. While the thin veil of emission across the shadow is caused by the dominant torus for low $R_{\rm high}$, it traces the jet foot-point for higher values of $R_{\rm high}$. Since the electron temperature in the torus is decreased significantly in the jet onset-dominated images, the veil cannot be identified with a plasma feature in the torus anymore, but has to be located further away from the shadow. 

\subsubsection{MAD simulation}

At $R_{\rm high}=1$, the source morphology is comparable to the corresponding SANE case for both spacetimes since the emission from the disk is dominating and the electron temperature is similar in all cases, as we can observe in panels b and f. The overall source size is smaller in the dilaton system. This is consistent with Fig. \ref{fig:SBG2Te_MAD}, where panel a shows that in the MAD case the $\sigma=1$ contour line traces a much larger jet opening angle than it does in SANE (panel a in Fig. \ref{fig:SBG2Te_SANE}).  

Moving to $R_{\rm high}=10$, the MAD images for both spacetimes reach a source morphology that remains unchanged for $R_{\rm high}>10$. Kerr-- and dilaton images show a very similar source structure, due to the distribution of electron temperature outside $\sigma=1$, which in MAD is more similar between spacetimes compared to the SANE case (compare panels c and d of Fig. \ref{fig:SBG2Te_MAD}). Albeit much smaller, the main difference apart from the shadow size is again the appearance of the receding side. In MAD images, it is almost as faint in the dilaton images as it is in the Kerr case (e.g. panels a an d on the right side in Fig. \ref{fig:GRRT_ISCO}. 

In $R_{\rm high}\geq10$ images, the electron temperature in the torus is decreased and a thin arc spans across the shadow, tracing the jet foot-point. The jet base in the MAD simulation is wider than in the SANE case; this is explained by comparing panels a and d for the Kerr or for the dilaton black hole in Figs. \ref{fig:SBG2Te_MAD} and \ref{fig:SBG2Te_SANE} (see Sec. \ref{sec:grmhd_results}). The region of peak emission is mostly confined to a $\sim20\,\upmu$as\,$\times\,20\,\upmu$as patch located at the top left of the shadow.  
Around $30\%$ of total emission is concentrated in the left half in all MAD images; while  this was the case for the Kerr black hole in SANE at least for $R_{\rm high}\leq10$, for the dilaton system this is a stark contrast to the SANE simulation. 

\subsubsection{Image comparison}

\begin{figure}
	\centering
	\includegraphics[width=0.5\textwidth]{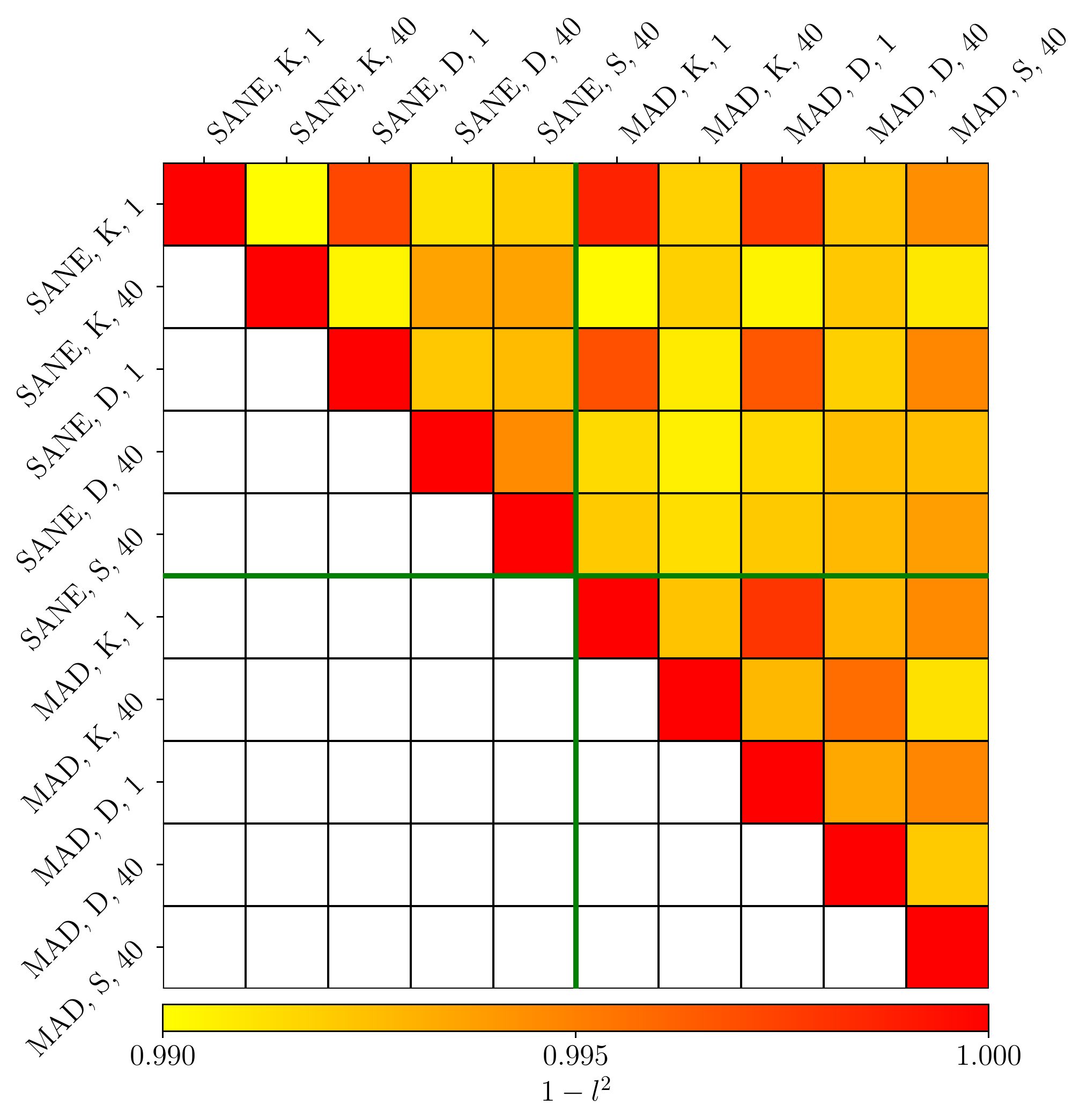}
	\caption{ $\rm L_{2}$ norm of the pixel-by-pixel differences between images of given models. The upper left and lower right quadrants show comparisons of spacetime-$R_{\rm high}$ combinations in SANE and MAD, respectively. The upper right quadrant shows comparisons also for different accretion models. The labels are abbreviated as "K": Kerr, "D": dilaton, "S": Schwarzschild, and $R_{\rm high}=1,40$.} \label{fig:carpet}
\end{figure}

In order to gain a wider overview of the differences between the various models, we compute the $\rm L_{2}$ norm of the pixel-by-pixel differences between images of a given electron distribution function, but varying spacetime, accretion model and $R_{\rm high}$ parameter in the electron temperature model. Figure \ref{fig:carpet} depicts $\rm l_{2}=1- L_{2}$, in such a way that on the diagonal the comparison between identical models yields the identical norm value $l_{2}=1$ (red fields). Since the plot is symmetric, we only show the upper triangle for clarity. Throughout the parameter space, a common feature is that $R_{\rm high}=1$ images show a high degree of similarity, which is consistent with plasma properties known from GRMHD. Overall, comparisons of SANE models show larger differences (yellow fields) to other SANE models than comparisons within the MAD parameter space. Generally, the largest differences appear for combinations with different $R_{\rm high}$. Comparing the upper left and lower right quadrants to the upper right one, it is evident that the differences among combinations of models with mixed spacetime, accretion model and electron temperature are not clearly distinguishable from comparisons between a Kerr and a dilaton model with fixed accretion and emission model, and either a thermal or a non-thermal emission model.

\subsection{Spectral analysis}

\begin{figure*}
	\centering
	\includegraphics[width=\textwidth]{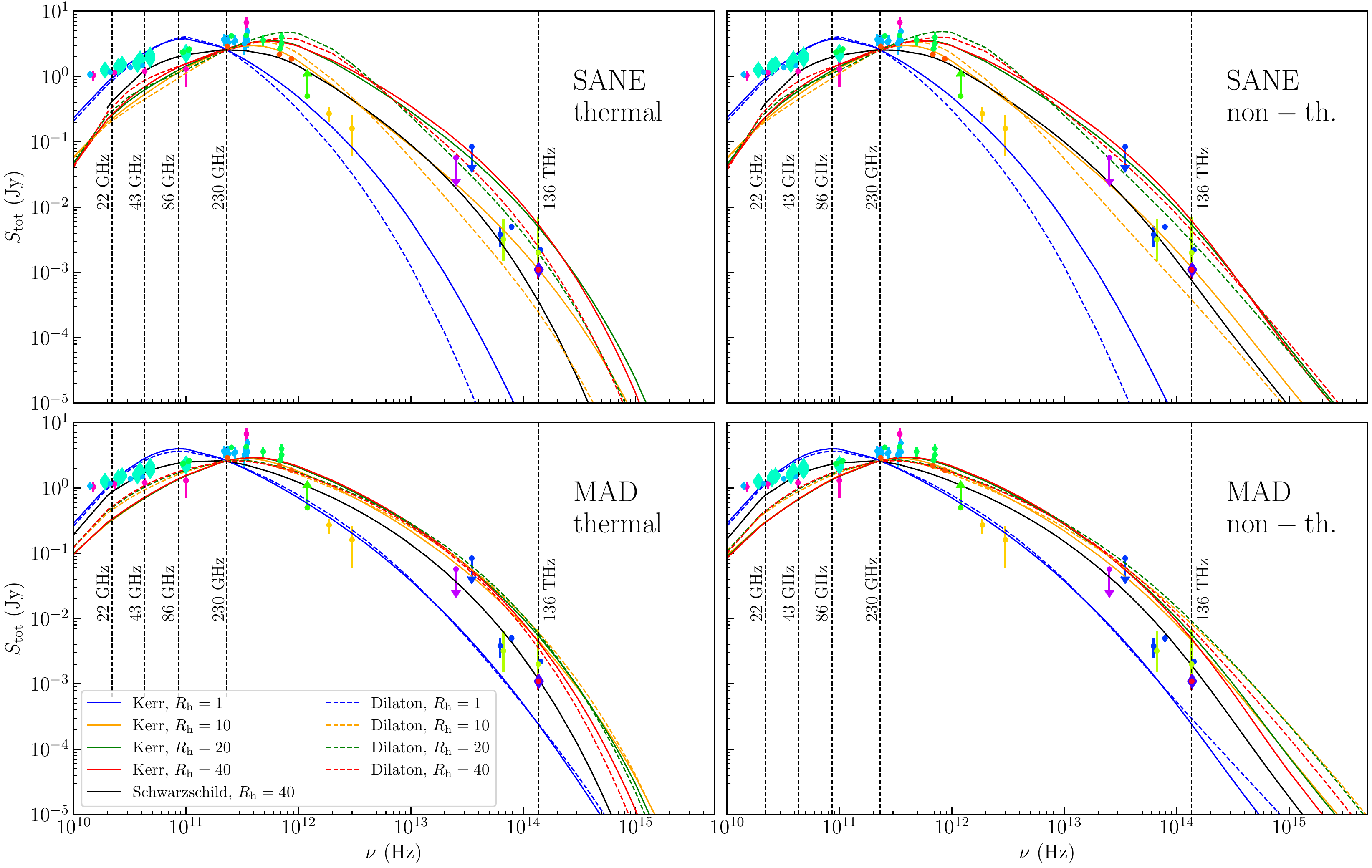}
	\caption{Time-averaged spectral energy distributions for SANE/MAD and thermal/non-thermal models. For all non-thermal models, $\varepsilon=0.015$. Over-plotted: observational data (see Table \ref{tab:obsSED}).} 
	\label{fig:SED_SgrA}
\end{figure*}

Figure \ref{fig:SED_SgrA} shows thermal and non-thermal model broad-band spectra for SANE and MAD cases, together with observational total flux measurements (see Table \ref{tab:obsSED} for details). Regardless of background spacetime, accretion model and eDF, the $R_{\rm high}=1$ spectrum behaves strikingly different compared to all other models due to the "cooling" effect of non-thermal particles in the jet introduced by the R-$\beta$ model, i.e. we see purely thermal emission from the accretion disk. While they appear to be the best fit to observational data for frequencies below the $R_{\rm high}=1$ turnover of about 80\,GHz, determined by the low magnetic field in the accretion disk (see Eq. \ref{eq:turnoverfreq}), they deviate greatly from observations for $\nu\geq230$\,GHz  and are therefore excluded from further discussion. 

To complete our comparison, we include Schwarzschild simulations for $R_{\rm high}=40$. In SANE, the corresponding SEDs are comparable to $R_{\rm high}=10$ SEDs of both Kerr and dilaton spacetimes, while in MAD they lie distinctly in between $R_{\rm high}=1$ and $R_{\rm high}>1$ SEDs of Kerr and dilaton curves. For a more detailed discussion on the Schwarzschild simulations, see Appendix \ref{sec:schwarzschild}.

Before comparing the Kerr to the dilaton spectra, it is important to
understand the general behavior of the Spectral Energy Distribution (SED)
under changes of $R_{\rm high}$ and the eDF. In the weakly magnetized
disk, high values of plasma-$\beta$ lead to an inverse dependence of the
electron temperature on $R_{\rm high}$, i. e. $\Theta_{\rm e}\sim R_{\rm
  high}^{-1}$ (see Eqs. \ref{eq:tratio_beta} and
\ref{eq:dim_less_T_e}). Applying a kappa eDF in the jet sheath as
described above, an additional high energy contribution in the form of a
power law tail is introduced in the SED, in contrast to the exponential
decay of the thermal models. If we now use the aforementioned dependence
of the electron temperature on $R_{\rm high}$ in the expressions for
thermal and kappa synchrotron emissivity $j_{\nu,\rm tot}$ taken from
\cite{Pandya2016}, we can write the high energy part of the SED as
\citep{Fromm2021b}

\begin{equation}
j_{\nu,\rm tot}\propto\exp\left(-R_{\rm high}^{2/3}\nu^{1/3}\right)+\nu^{-(\kappa-2)/2} \left[1/R_{\rm high} + \varepsilon\sigma\right]^{\kappa-2},
\label{eq:jnukapprox}
\end{equation}
where $\sigma$ is again the magnetization and $\varepsilon$ sets the magnetic contribution to the energy of the electrons. The first term in the above equation describes the thermal contribution to the total emission and the steepening of the high energy part of the SED with increasing $R_{\rm high}$ (most prominent in the near infrared, $\nu\gtrsim1.36\times10^{14}$\,Hz). On the other hand, the second term adds the non-thermal electrons in the jet wall and decreases the dependence on $R_{\rm high}$ of the high frequency ($\nu\gtrsim2\times10^5$\,GHz) spectrum compared to the purely thermal case (see top row of Fig. \ref{fig:SED_SgrA}). The lower energy part of the SED is governed by the jet, characterized by low to intermediate plasma-$\beta$,  where the electron temperature is effectively independent of $R_{\rm high}$.

Lastly, the turnover position is affected by the choice of $R_{\rm high}$, as well as mass accretion rate and magnetic field strength 
\citep[see e.g.][and \citealt{Fromm2021b} for details]{Zdziarski1998}.  
\begin{linenomath}\begin{equation}
\nu_{\rm t,th}\propto B_{\rm code}\sqrt{\dot{m}}/R_{\rm high}^2.
\label{eq:turnoverfreq}
\end{equation}\end{linenomath}
The above equation explains the shifts in turnover with increasing $R_{\rm high}$ in the SANE models, and the reduced shift in the MAD cases where the magnetic field is much stronger and the mass accretion rate is smaller.
In SANE, for $\nu\leq230$GHz the total flux slightly increases with $R_{\rm high}$ (top left panel of figure \ref{fig:SED_SgrA}). As explained above, an increase in $R_{\rm high}$ decreases the electron temperature and subsequently the emission from the disk, making the jet comparatively brighter when fixing the total 230\,GHz flux of the average image to the same value. Regardless of $R_{\rm high}$ and eDF, the dilaton SEDs turn over at higher frequencies with higher total fluxes, and the difference in turnover position between spacetimes increases with $R_{\rm high}$. For example, for a thermal distribution, at $R_{\rm high}=10$ the turnover happens at $\sim4\times10^{11}$\,Hz for Kerr and $\sim5\times10^{11}$\,Hz for the dilaton case, whereas for $R_{\rm high}=40$ the turnovers move to $\sim5.5\times10^{11}$\,Hz and $\sim7.5\times10^{11}$\,Hz, respectively. Past the turnover, in the thermal case dilaton SEDs remain steeper than their Kerr counterparts throughout the rest of the frequency range. Above a certain frequency between $2\times10^{12}$ and $8\times10^{12}$\,Hz, the total flux of the dilaton models drops below the corresponding Kerr SEDs. In the near infrared, around $1.36\times10^{14}$\,Hz, the differences in total flux between spacetimes can be of an order of magnitude.  

\begin{figure*}
	\centering
	\includegraphics[width=0.45\textwidth]{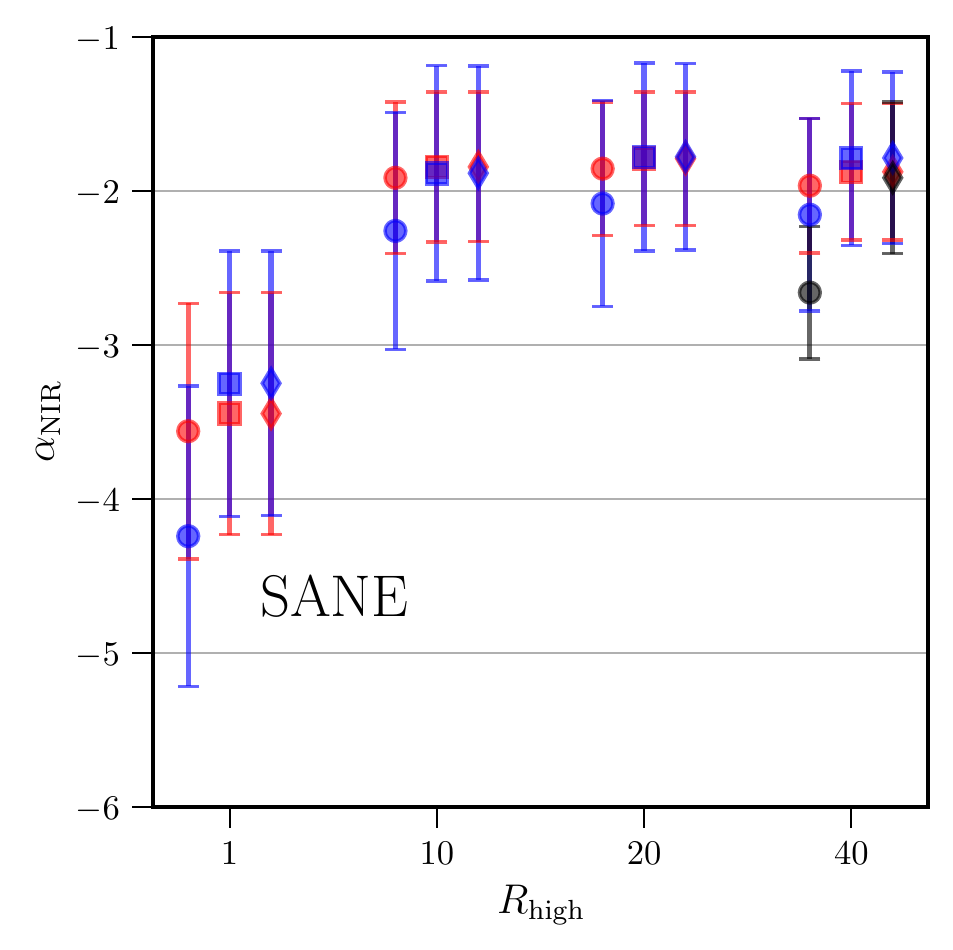}
	\includegraphics[width=0.45\textwidth]{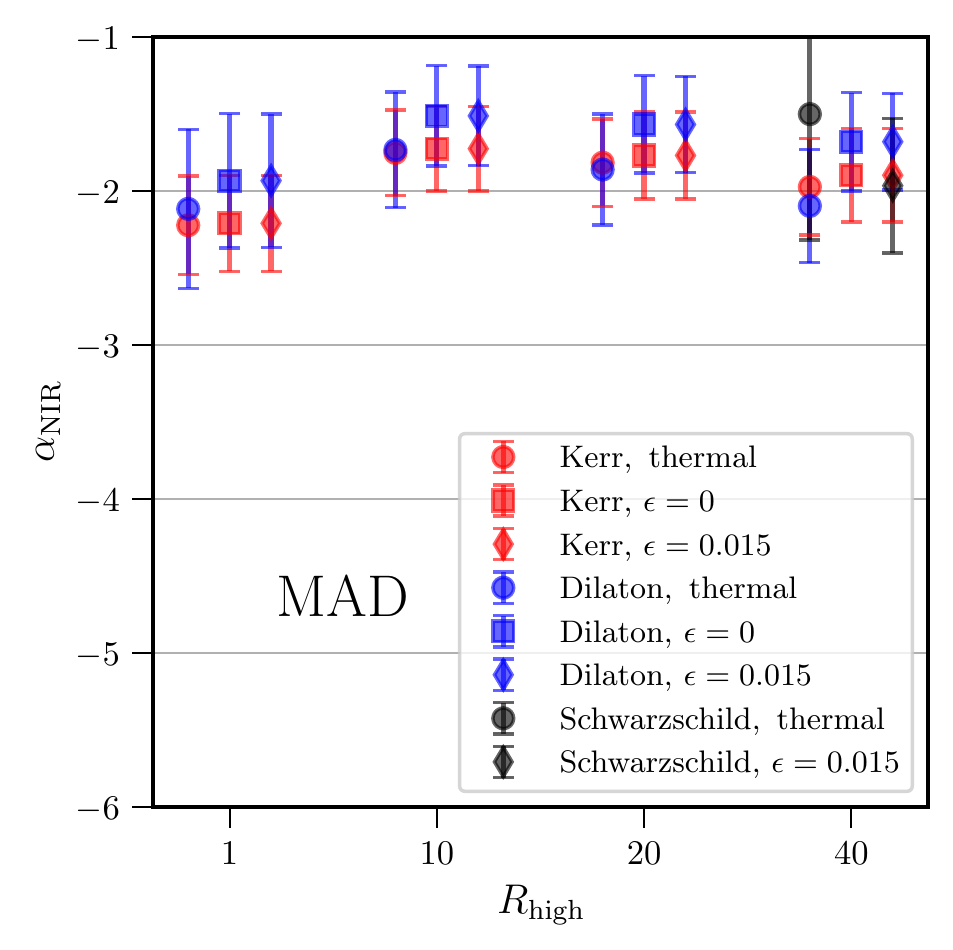}
	\caption{Near infrared spectral indices for the SANE (left panel) and MAD (right panel) simulations, obtained from the time-averaged spectra (Fig. \ref{fig:SED_SgrA}).} 
	\label{fig:alphaNIR_MAD}
\end{figure*}

Non-thermal emission, despite being solely applied in the jet, introduces the characteristic power-law tail in the near infrared (NIR) for all $R_{\rm high}\geq10$ SEDs for $\nu\gtrsim10^{14}$\,Hz \citep[see][]{Cruz-Osorio2021b}. For a given spacetime, the already small dependence on $R_{\rm high}$ for $R_{\rm high}\geq20$ is decreased even further in the NIR. The flattening due to the non-thermal contribution is slightly more pronounced for the dilaton SEDs, so that for $R_{\rm high}\geq20$ Kerr and dilaton SEDs lie closely together, and for $R_{\rm high}\geq20$ the Kerr SEDs are steeper past $1.36\times10^{14}$\,Hz.  Figure \ref{fig:alphaNIR_MAD} visualizes the above observation: thermal SEDs show NIR spectral indices $\alpha_{\rm NIR}\sim-2.3$, while non-thermal models are flatter in the $\alpha_{\rm NIR}\sim-2.0$, consistently for $R_{\rm high}\geq10$ and either spacetime. Thermal dilaton SEDs are steeper, and non-thermal ones are slightly flatter compared to their Kerr counterparts, respectively. The $R_{\rm high}=10$ curves are separated by more than an order of magnitude from the other SEDs, and show the same separation between Kerr and dilaton SEDs (yellow solid and dashed lines in the top right panel).\\

In thermal MAD SEDs, the dependence on both $R_{\rm high}$ and the background spacetime is much weaker compared to the SANE case, especially between the turnover positions and the NIR (bottom left panel in Fig. \ref{fig:SED_SgrA}). The turnovers of Kerr-- and dilaton SEDs are much closer together and show much more similar total fluxes in the MAD models. With increasing $R_{\rm high}$, the high energy part shows the expected steepening. In contrast to the SANE models, the $R_{\rm high}=10$ curves are no longer clearly separated from the other SEDs. Introducing non-thermal emission, dilaton SEDs again flatten more than the respective Kerr models, leading to a more clear separation of SEDs of different spacetimes (bottom right panel in Fig. \ref{fig:SED_SgrA}). The NIR spectral index is plotted in Fig. \ref{fig:alphaNIR_MAD}, indicating steeper thermal dilaton SEDs (compared to Kerr), but much flatter non-thermal SEDs. This behavior, albeit weaker, is also present in the SANE models as described above.\\

The observed NIR flux and spectral index of Sgr\,A$^\ast$ is highly variable \citep[e.\,g.][]{Witzel2014,Witzel2018}. In a bright or flare state, $\alpha_{\rm NIR}=-0.6\pm0.2$ was determined from synchronous observations at $8.102\times10^{13}$\,Hz and $1.874\times10^{14}$\,Hz \citep{Hornstein2007,Witzel2014}. The NIR spectral indices calculated from the Kerr and dilaton spectra (Fig. \ref{fig:alphaNIR_MAD}) are clearly inconsistent with a such a steep value. They indicate a quiescent state of the systems, regardless of background spacetime, accretion model or emission model (universally $\alpha_{\rm NIR}\lesssim-1.50$). The spectral indices are consistent with dim state measurements giving $\alpha_{\rm NIR}=-1.7\pm-0.4$ \citep{Gillessen2006} and $\alpha_{\rm NIR}=-1.64\pm0.06$ \citep{Witzel2018}, or even steeper values \citep[see e.g.][for details]{Witzel2014}.
In terms of total NIR flux, the SANE SEDs fit the observational data better than the MAD ones. More precisely, the top row of Fig. \ref{fig:SED_SgrA} shows that Kerr $R_{\rm high}=10$ images for either electron distribution function well match the $1.36\times10^{14}$\,Hz flux reported by \cite{Gravity2020c} (bright pink point on the 136\,THz line indicated in each plot). Among the thermal SEDs, $R_{\rm high}\geq20$ dilaton SEDs not only match various $1.36\times10^{14}$\,Hz measurements, but also those taken around the $3.0\times10^{13}$\,Hz mark. The corresponding non-thermal dilaton SEDs tend to slightly overshoot the NIR observations. While around $3.0\times10^{13}$\,Hz the MAD SEDs match the data well, they collectively overshoot the NIR flux. 

\section{Summary and Discussion}\label{sec:summary}

In this work, we have investigated the possibility to distinguish between different spacetimes by means of simulations and observations of black holes, under the aspect of different emission models. To this end, we first carried out SANE and MAD GRMHD simulations of a Kerr and a dilaton black hole in full 3D. In radiative post-processing, we first parametrized the electron temperature and studied the effect of the R$-\beta$ parametrization on GRRT images obtained with a purely thermal electron energy distribution. We subsequently repeated this process with a non-thermal kappa electron distribution applied in the jet wall, with an optional contribution of magnetic energy. For each model, we fixed the mass accretion rate to fit the flux of an average GRRT image to 2.5\,Jy at 230\,GHz. Further, we computed synchrotron SEDs and near-infrared spectral indices for all models. For each emission model, we computed image comparison metrics between Kerr and dilaton images to quantify differences. 
\subsection{GRMHD simulations}

The goal of this work is a theoretical comparison of two background spacetimes, with the comparison to observational data playing only a minor role. Therefore, the spin of the Kerr black hole is fixed to $a=0.6$, and in the process the dilaton parameter is fixed to $\hat{b}=0.5$ by matching the black holes at their ISCO. Once both systems have entered a quasi steady state in their evolution (past 10,000\,M), the Kerr black hole shows wider jet opening angles in both SANE and MAD simulations compared to the dilaton spacetime, as well as a higher magnetized jet. Overall, the two systems behave rather similarly in either accretion model. In the MAD simulations, neither systems fully reaches the "MAD state" characterized by $\psi>10$ \cite{Tchekhovskoy2011}.

\subsection{GRRT calculations}

\subsubsection{Spectral analysis}
From the multi-wavelength GRRT calculations, we generate time-averaged broad-band spectra and 230\,GHz images. We employ the R$-\beta$ parametrization, choosing $R_{\rm high}\in{[1,10,20,40]}$. In the next step, we apply a non-thermal kappa electron energy distribution function in the jet wall, with and without an additional magnetic contribution described by $\varepsilon=0$ and 0.015, respectively. In the spectra, the accretion model affects the position of the turnover point, and non-thermal emission introduces a power-law tail in the near infrared compared to the steep decrease in the thermal case. The dependence of the spectra of a given emission model on the background spacetime is much more prominent in the SANE case, where for a given frequency the total flux can differ by an order of magnitude between spacetimes. In the thermal case, an increase of $R_{\rm high}$ steepens the high energy part of the spectrum; with non-thermal emission, for $R_{\rm high}\geq20$ also the dependence on the spacetime decreases significantly. Up until the near-infrared, the MAD spectra are almost independent of both spacetime and emission model. In the non-thermal case, the NIR spectral indices of Kerr- and dilaton black hole systems are $\gtrsim0.25$ apart. For the thermal models, there is a clear difference only for $R_{\rm high}\geq20$ (see Fig. \ref{fig:alphaNIR_MAD}). While the reported differences in spectral indices are potentially observable features, the spectra are universally steeper than recent observations indicate \citep{Witzel2018}. Observed NIR spectral indices of Sgr\,A$^\ast$ range from -0.6 to -1.64 \citep{Witzel2018}.
In order to better match observations, the target flux of the average image and the contribution of magnetic energy of the kappa electron distribution may be increased in a follow-up study. Likewise, $\sigma_{\rm cut}$ can be increased to modify the extent and position of the jet sheath. 

\subsubsection{GRRT images}

The R$-\beta$ parametrization splits the source morphology into torus- ($R_{\rm high}\lesssim10$) and jet dominated ($R_{\rm high}\gtrsim20$) images. In SANE, the transition is rather smooth and takes place between $R_{\rm high}=10$ and 20, for both Kerr- and dilaton black holes. The wider jet opening angle in the Kerr system indicated in the GRMHD simulations translates to the jet dominated images, and due to increased Doppler boosting from the rotation of the Kerr black hole the receding side is particularly faint. Since the dilaton black hole in non-rotating, the receding side is more prominent. The filamentary low flux features are more fuzzy compared to the Kerr images, and the shadow is smaller due to the ISCO match. 
The transition from torus to jet dominated images is rather abrupt for the MAD models: before $R_{\rm high}=10$, the source morphology converges and stays the same for $R_{\rm high}\geq10$. The jet opening angle is more similar between spacetimes in the MAD images, and apart from the shadow size the source morphologies of Kerr and dilaton systems are visually identical. 

\subsection{Differently matched black holes}\label{seq:matchings}

This study focused on the special case of Kerr and dilaton black holes matched at their ISCO, concluding that distinguishing between spacetimes is still challenging even in a non-observational framework. To reinforce our argument, we consider SANE simulations of the dilaton black hole matched to the Kerr one at its unstable circular photon orbit, and at its event horizon (see Appendix \ref{sec:matchings_app}). As for the ISCO case, the dilaton characteristic radii are always matched to the equatorial equivalent in the Kerr spacetime. We summarize below the size of the matched radii and their apparent size on the sky, scaled to the mass and distance of Sgr\,A$^\ast$.
\begin{linenomath}\begin{align}
&{\rm I.} & \ \ r_{\rm ISCO}^{\rm dilaton}&=r_{\rm ISCO,\ equatorial}^{\rm Kerr}&\approx 3.83\,M &\approx 19.17\, {\rm \upmu as} \\
&{\rm II.} & \ \ r_{\rm PH}^{\rm dilaton}&=r_{\rm PH,\ equatorial}^{\rm Kerr}&\approx 2.19\,M &\approx 10.96\, {\rm \upmu as}\\
&{\rm III.} & \ \ r_{\rm EH}^{\rm dilaton}&=r_{\rm EH,\ equatorial}^{\rm Kerr}&\approx 1.80\,M &\approx 9.01\, {\rm \upmu as}
\end{align}\end{linenomath}
From Figs. \ref{fig:GRRT_ISCO} and \ref{fig:GRRT_PO_EH} it is evident that moving away from a match at the ISCO, the ability to distinguish between spacetimes decreases further. For photon orbit and event horizon matches, the dilaton source morphology becomes more similar to that in the Kerr system for any value of $R_{\rm high}$.
Even if the images differ in terms of plasma features, those may as well have astrophysical causes and need not be gravitational (that is, looking at two images without prior knowledge of the background metrics). Comparing Figs. \ref{fig:GRRT_ISCO} and \ref{fig:GRRT_PO_EH} it is apparent that choosing a different matching case on the one hand has a minuscule effect on the visual appearance of the shadow size, and on the other hand affects the source morphology so that the dilaton system looks progressively more similar to the Kerr system moving to smaller characteristic radii for the match. In the photon orbit and event horizon matchings, the spacetimes are hence even harder to distinguish. 

\subsection{Limitations of the models}
In this study, we investigate only a small fraction of the available parameter space. 
Comparisons to rotating dilaton black holes could be another valuable addition to this study. Further, magnetic field geometries and initial conditions of the torus can affect the evolution of the black hole system \citep{Cruz2020}. Finally, our simulations only evolve the dynamically important protons, thereby neglecting effects of electron heating mechanisms \citep[e.\,g.][]{Chael2018,Mizuno2021}, radiative feedback and resistivity of the plasma \citep[see, e.\,g.][]{Ripperda2020}.

In the GRRT calculations, the R$-\beta$ parametrization emulates electron heating processes in the vicinity of black holes \citep{Mizuno2021}. Alternative prescriptions for the electron temperature \citep{Anantua2020} could alter the source structure considerably. When employing the kappa electron distribution function, the inner boundary of the jet wall can be modified to change the size of the region containing accelerated electrons. In the same vein, the magnetic contribution to the distribution function could enable us to better match observed NIR spectral indices. The inclination can further enhance the prominence of the jet in the GRRT images. Lastly, including polarization in the GRRT calculations would enable us to map the magnetic field geometry in the images. These combined effects and extensions to the study could increase the chances of distinguishing between two spacetimes.
This work is a phenomenological approach to the goal of testing general relativity in an imaging framework comparing exemplary models. A model-independent approach through feature extraction from the images, such as fitting crescent or ring models to images and visibilities and analyzing emission profiles, could help us to better quantify differences between spacetimes.

\section{Conclusion}\label{sec:conclusion}

Combining the results from GRMHD and GRRT simulations, we conclude that it is still challenging to distinguish black holes characterized by different background metrics, at least in the case of the dilaton metric. The overall behavior of the GRMHD simulations is very similar in the MAD case, even more so than in the SANE simulations, due to the matching at the ISCO. From the GRRT images, we see that the accretion and emission models have a much larger impact on the source morphology than the underlying spacetime does. The $R_{\rm high}$ parameter alone changes the source morphology drastically from torus to jet dominated; this transition is smooth in SANE, but takes place abruptly in MAD for some $R_{\rm high}<10$. The prominent, potentially observable differences between spacetimes in the GRRT images can be summarized as follows:
\begin{itemize}
	\item The jet opening angle is wider in the Kerr spacetime;
	\item The receding side of the torus is fainter in Kerr due to increased Doppler boosting;
	\item The Kerr shadow is larger than the dilaton shadow due to the ISCO match.
\end{itemize}
From the spectra, the differences between spacetimes in near-infrared spectral index and total flux potentially suffer from degeneracy between accretion model, emission model and spacetime. It is questionable whether even fitting the whole spectrum to observational data would enable us to distinguish between spacetimes. 

Including a Schwarzschild black hole in our investigation shows that the differences in image space to the Kerr and dilaton black hole are larger than between the latter two spacetimes, but overall remain small. From the comparison metrics (see Fig. \ref{fig:carpet}), the Schwarzschild metric is indistinguishable from the other considered models.

\begin{acknowledgements}
We thank Dr. G. Witzel and Dr. N. MacDonald for their role as internal referees at the MPIfR and for helpful discussions and comments, as well as Dr. P. Kocherlakota for his perspective and comments.
JR received financial support for this research from the International Max Planck Research School (IMPRS) for Astronomy and Astrophysics at the Universities of Bonn and Cologne. This research is supported by  the European Research Council for advanced grant ``JETSET: Launching, propagation and emission of relativistic jets from binary mergers and across mass scales'' (Grant No. 884631) and European Horizon Europe staff exchange (SE)
programme HORIZON-MSCA-2021-SE-01 Grant No. NewFunFiCO-101086251.
CMF is supported by the DFG research grant ``Jet physics on horizon scales and beyond'' (Grant No. FR 4069/2-1). ZY is supported by a UKRI Stephen Hawking Fellowship and  acknowledges support from a Leverhulme Trust Early Career Fellowship. YM acknowledges the support by the National Natural Science Foundation of China (Grant No. 12273022). The simulations were performed on GOETHE-HLR LOEWE at the CSC-Frankfurt, Iboga at ITP Frankfurt, and SuperMUC-NG in Garching. 
\end{acknowledgements}

{\noindent\tiny {\it Software.} {\tt BHAC}\footnote{\href{https://bhac.science/}{https://bhac.science/}} \citep{Porth2017}, {\tt BHOSS} \citep{Younsi2020}}

\bibliographystyle{aa}
\bibliography{aeireferences_use}

\begin{appendix}

\section{Matching spacetimes}\label{sec:matchings_app}

\begin{table*}
	\centering 
	\def\arraystretch{1.5}
	\caption{Characteristic radii}
	\begin{adjustbox}{max width=0.99\textwidth}
	\begin{tabular}{lll}
		\hline\hline
 Kerr& Dilaton & Match \\
		\hline
	 $r_{\rm EH}=M+\sqrt{M^2-a^2}$ & $r_{\rm EH}=2\left(M-\hat{b}\right)$  & $\hat{b}_{\rm EH} =\frac{1}{2}\left(M-\sqrt{M^2-a^2}\right)$ \\ 
	 $r_{\rm PS}=2M\left(1+\cos\left[\frac{2}{3}\arccos\left(\frac{a}{M}\right)\right]\right)$ & $r_{\rm PS}=\frac{1}{2} \left(3\left(M-b\right)+\sqrt{\left(M-b\right)\left(9M-b\right)}\right)$  & $\hat{b}_{\rm PS} =\frac{1}{2}M\left(-2-3C+\sqrt{8+C(C+8)}\right)$ \\
	 $r_{\rm ISCO}=M\left(3+Z_2+\sqrt{(3-Z_1)(3+Z_1+2Z_2)}\right)$ & $r_{\rm ISCO}=2M\left(B+B^2+B^3\right)$ & $\hat{b}_{\rm ISCO} = M\left(1+\frac{1}{27}\left[1+\sigma-\frac{\sigma}{2}\right]^3\right)  $ \\
		\hline
	\end{tabular}
	\label{tab:matching}
    \end{adjustbox}
	\tablefoot{Kerr and dilaton balck hole event horizon (EH) and equatorial photon sphere (PS) and innermost stable circular orbit (ISCO). The right column lists the dilaton parameter $\hat{b}$ in relation the the Kerr spin $a$ required to have the two black hole systems matched at the repsective radius. Parameters $B$, $C$, $Z_1$, $Z_2$ and $\sigma$ are reported in appendix \ref{sec:matchings_app}. All expressions are taken from the supplemental material to \cite{Mizuno2018}.}
\end{table*}

In order to study two comparable systems, they need to be matched at one of their characteristic radii. This can be the event horizon (EH), or the (equatorial) photon sphere (PS) or innermost stable circular orbit (ISCO). In this study, we mainly focus on the latter case to ensure similar plasma dynamics near the black hole. We match the spacetimes by simply equating the analytical expressions for the radii and arranging terms so that a spin in the Kerr system corresponds to a dilaton parameter in the other system. 

The dilaton spacetime in its analytic form reads \citep[taken from][]{Mizuno2018}
\begin{linenomath}\begin{equation}
    ds^2=-\left(\frac{r-2\mu}{r+2\hat{b}}\right)dt^2+\left(\frac{r+2\hat{b}}{r-2\mu}\right)d\rho^2+\left(r^2+2\hat{b}r\right) d\Omega^2,
\end{equation}\end{linenomath}
where $\rho^2=r^2+2\hat{b}r$ and $M=\mu+\hat{b}$ is the ADM mass. The metric is implemented in both GRMHD and GRRT codes in Rezzolla-Zhidenko parametrized form \citep{Rezzolla2014}. The corresponding coefficients are listed in Eqs. 10-16 in the supplemental material of \cite{Mizuno2018}.

The expressions for Kerr and dilaton characteristic radii are listed in table \ref{tab:matching}. The abbreviations used are reported below, where the subscript KI indicates the Kerr ISCO. $B$, $C$ and $\sigma$ are taken from \cite{Mizuno2018}.
\begin{linenomath}\begin{align}
    B=&\left(1-\frac{\hat{b}}{M}\right)^{1/3}\\
	C=&\cos\left(\frac{2}{3}\arccos\left[-\frac{a}{M}\right]\right)\\
	Z_1=&1+\sqrt[3]{1-\xi^2}\left(\sqrt[3]{1+\xi}+\sqrt[3]{1-\xi}\right), \ \ \xi=\frac{a}{M} \\
	Z_2=&\left(3\xi^2+Z_1^2\right)^{1/2}\\
	\sigma=&-\frac{7}{2} +\frac{3}{4M} \left(-9\,r_{\rm KI}  +\sqrt{36M^2+84M\, r_{\rm KI}+81\, r_{\rm KI}^2}  \right)
\end{align}\end{linenomath}

Figures \ref{fig:GRRT_ISCO} and \ref{fig:GRRT_PO_EH} show the GRRT images of Kerr and dilaton black holes matched at (equatorial) ISCO, unstable photon orbit and event horizon.

\begin{figure*}
	\centering
	\includegraphics[width=0.65\textwidth]{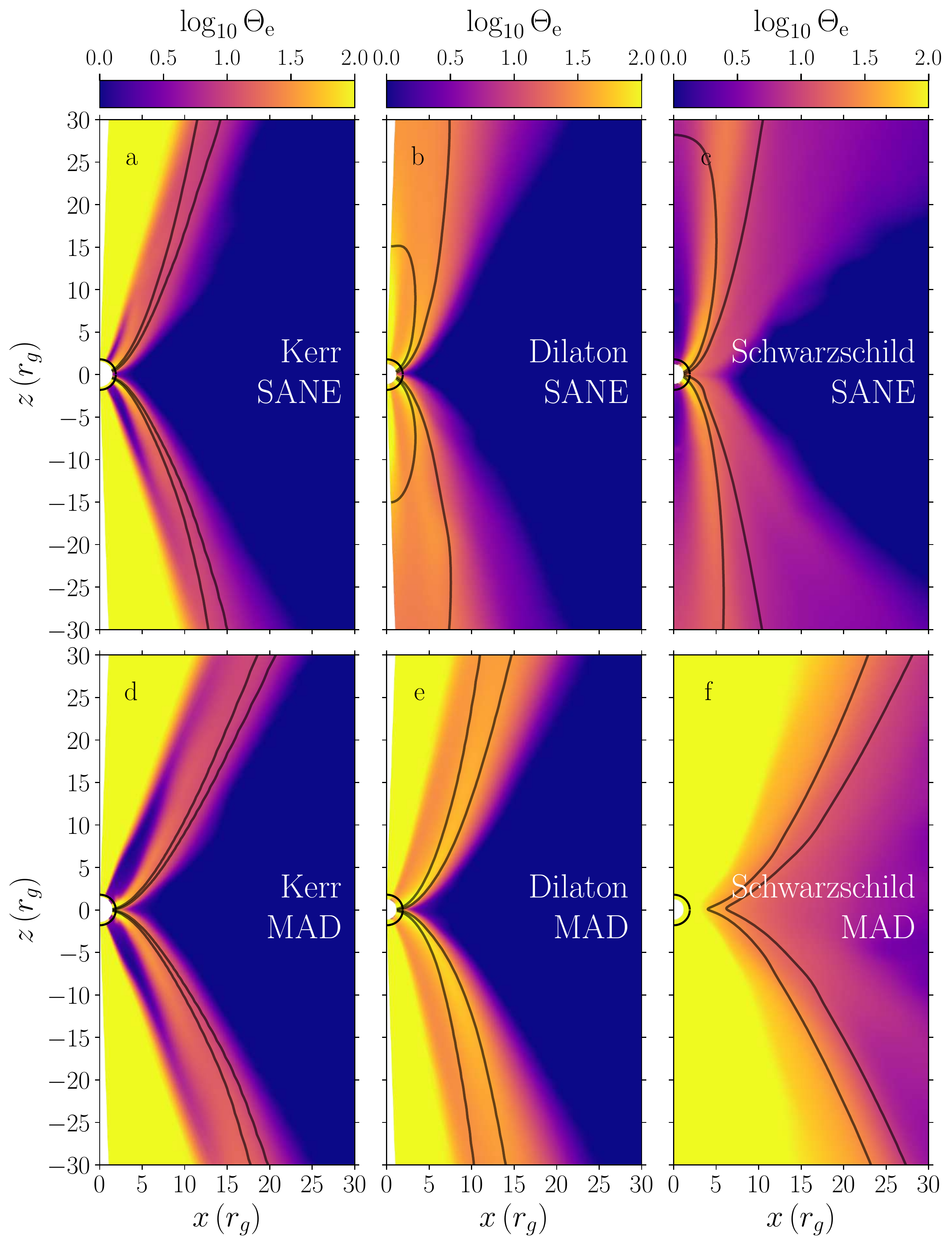}
	\caption{Electron temperature $\Theta_{\rm e}$ at $R_{\rm high}=40$ for SANE and MAD simulations in Kerr, dilaton and Schwarzschild spacetimes. Solid contour lines: levels of $\sigma$; line towards the pole $\sigma=1.0$, line towards the torus $\sigma=0.1$. The azimuthally averaged GRMHD data is shown time averaged over 1000\,M for Kerr and dilaton black holes, and 2000\,M for the Schwarzschild black hole.} 
	\label{fig:GRMHD_Schwarzschild}
\end{figure*}
\begin{figure*}
	\centering
	\includegraphics[width=0.495\textwidth]{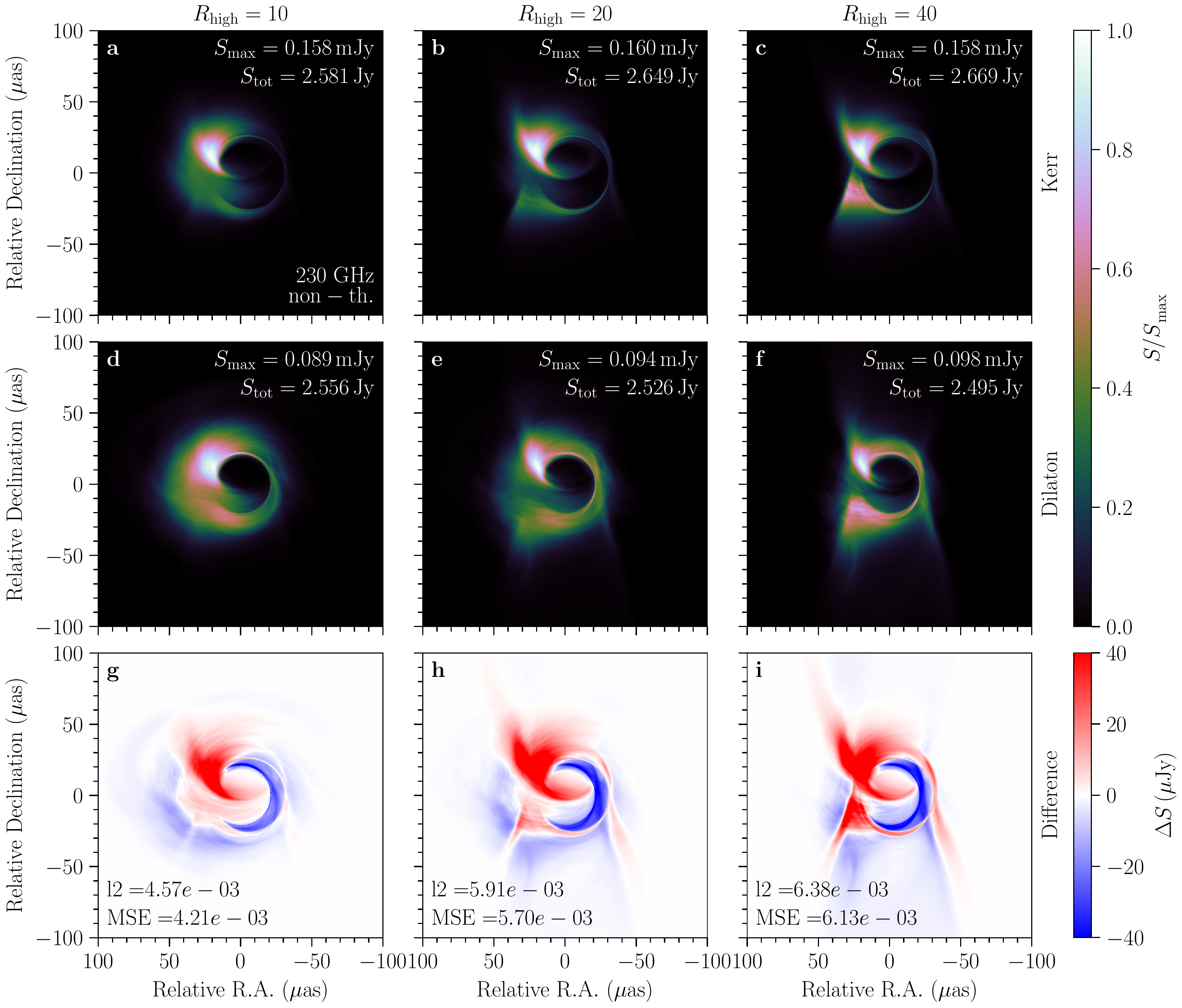}
	\includegraphics[width=0.495\textwidth]{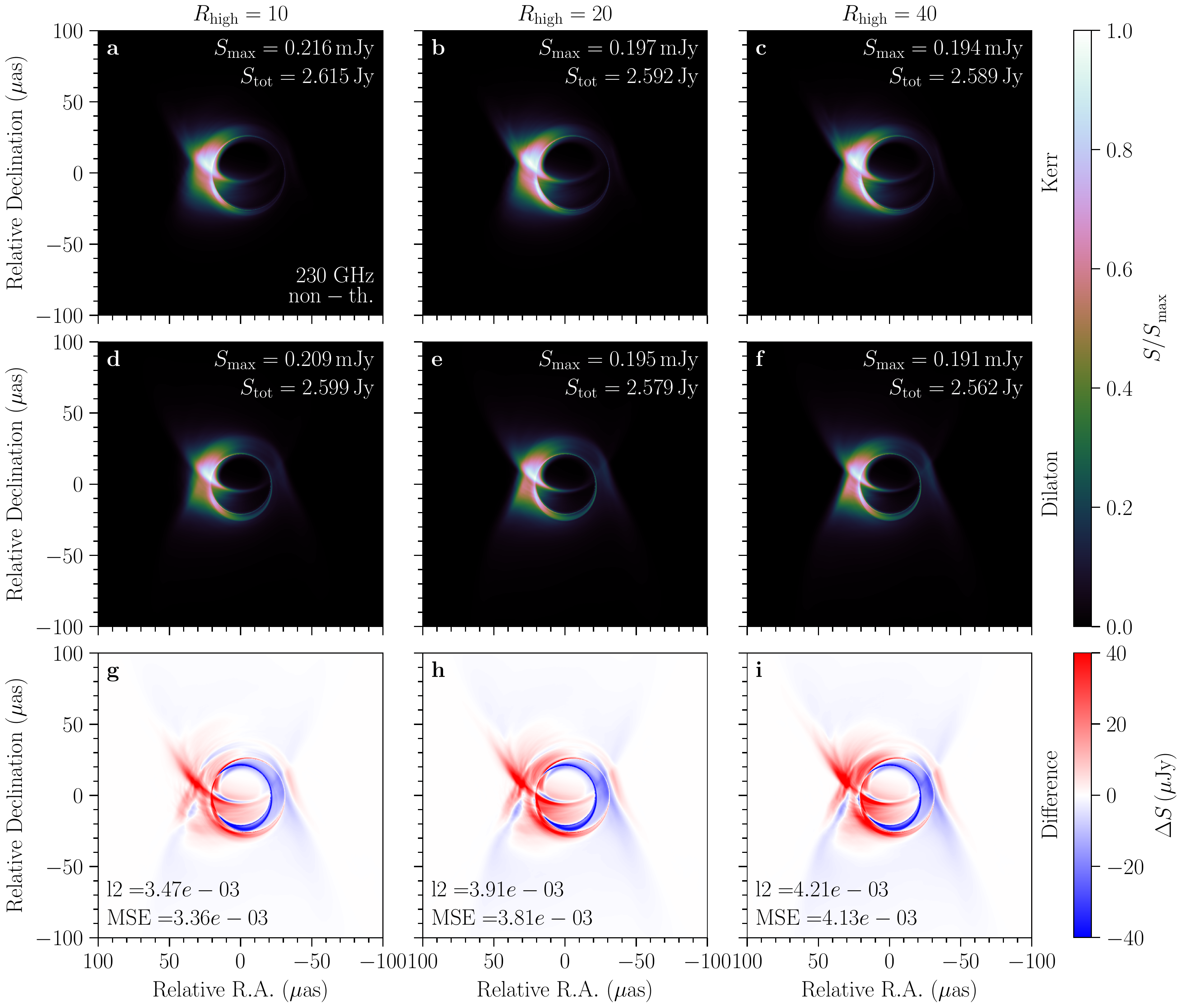}
	\caption{Kerr and dilaton GRRT images at $R_{\rm high}\in\{10,20,40\}$ in SANE (left) and MAD (right), matched at the ISCO for comparison to Fig. \ref{fig:GRRT_PO_EH}.
	The images are averages of 100 snapshots over 1\,000\,M simulation time ($\sim$$6$\,h for Sgr\,A$^*$). A non-thermal electron distribution function has been applied in the jet sheath.} 
	\label{fig:GRRT_ISCO}
\end{figure*}

\begin{figure*}
	\centering
	\includegraphics[width=0.495\textwidth]{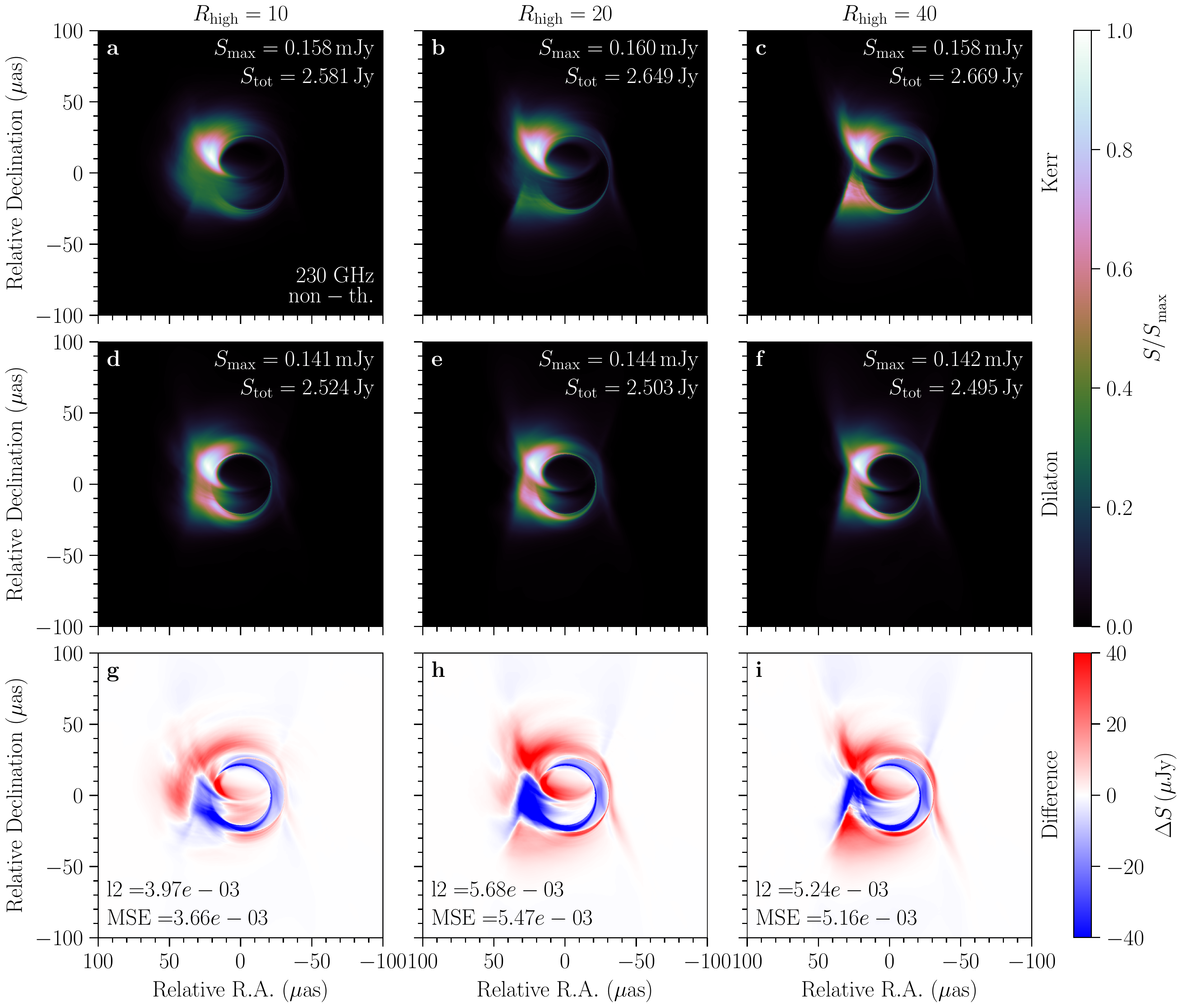}
	\includegraphics[width=0.495\textwidth]{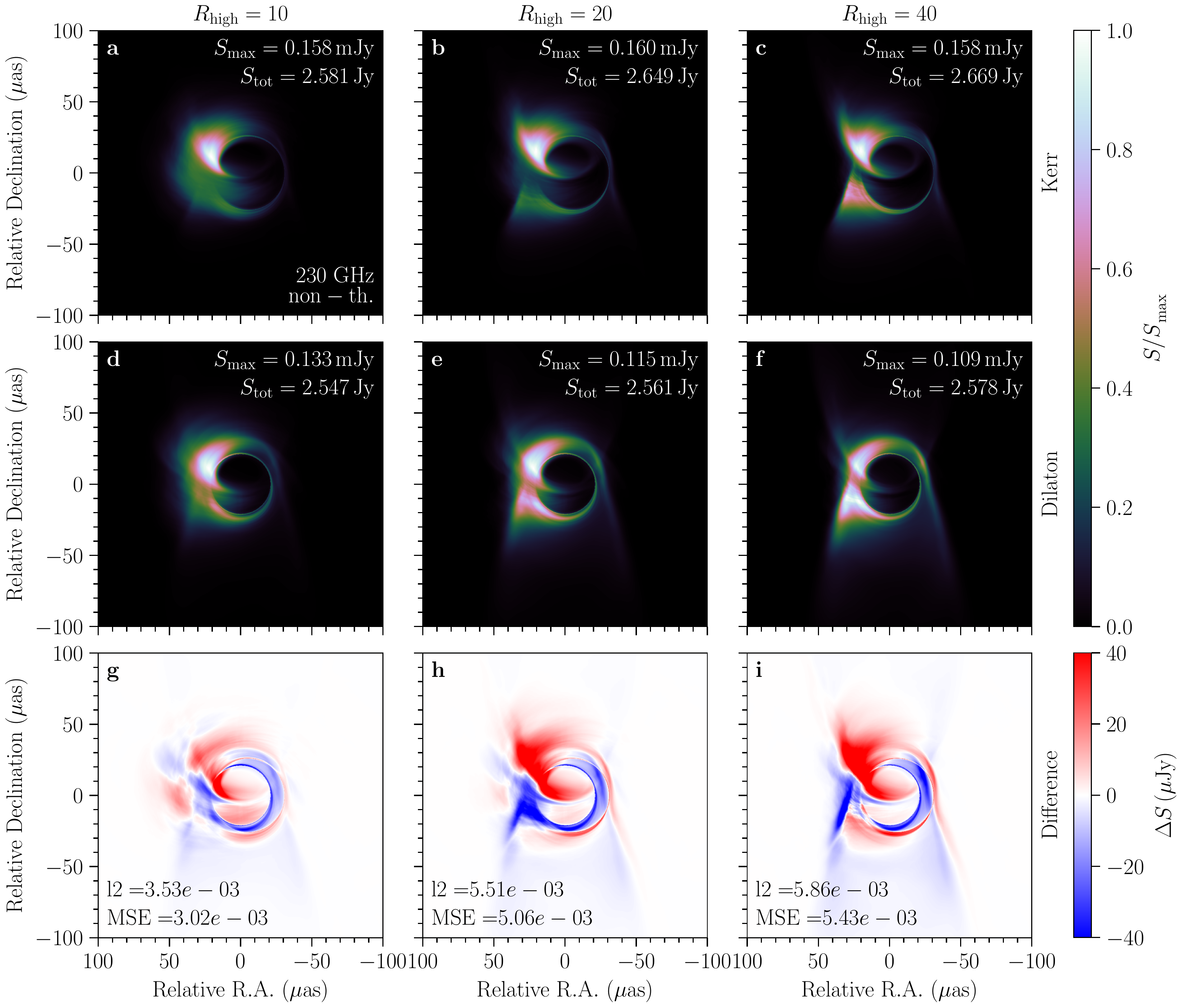}
	\caption{Same as Fig. \ref{fig:GRRT_ISCO}, but for the photon orbit and event horizon matched cases, in SANE.} 
    \label{fig:GRRT_PO_EH}
\end{figure*}

\section{Mass accretion rates}

During each GRRT calculation, a scaling parameter for the mass accretion rate is obtained to fix the total flux in the average image at 2.5\,Jy for 230\,GHz. Multiplied with the average GRMHD code accretion rate for the same time interval, one obtains the mass accretion rate in M$_\odot$/yr. Table \ref{tab:mdot_grrt} lists accretion rates all emission models. As explained in section \ref{sec:ET}, an increase in $R_{\rm high}$ decreases the electron temperature in-- and subsequently the emission from the torus. Therefore, a higher mass accretion rate is required to still match the specified target flux for a given electron distribution function (eDF). \\
At $R_{\rm high}=1$, differences between eDFs are nonexistent, and they remain rather small even when $R_{\rm high}$ is increased. Intuitively one could argue that for a kappa distribution, a lower accretion rate is required since the eDF naturally supplies more electrons with higher energies; however, the high-energy tail comes at the cost of an overall flatter distribution. This difference in lower-energy electrons overpowers those in the high-energy tail, requiring a slightly higher accretion rate to match the target flux.\\
Generally, SANE $R_{\rm high}=10$ models and MAD $R_{\rm high}\geq10$ models are in good agreement with the estimate of \cite{Bower2018} of $\dot{M}\sim10^{-8}$\,M$_\odot$/yr for Sgr\,A$^\ast$.
\vspace{-20pt}
\begin{table}[H]
	\noindent
	\centering 
	\def\arraystretch{1.5}
	\caption{Mass accretion rates for all Kerr and dilaton model configurations in thermal, kappa with $\varepsilon=0$ and kappa with $\varepsilon=0.015$, in units of 10$^{-8}$\,M$_\odot$/yr (Kerr $a_\star=0.6$, Dilaton $b=0.504$). $\dot{M}$ obtained by multiplying the BHOSS accretion parameter by the respective GRMHD accretion rates reported in table \ref{tab:grmhd_mean_stdev}. All $\dot{M}$ obtained to fit the 230\,GHz flux to 2.5\,Jy between 11\,000\,M and 12\,000\,M.}
	\begin{adjustbox}{max width=0.48\textwidth}
	\begin{tabular}{ l   l |llll}
		
		 \hline \hline 
		 \multirow{2}{1cm}{Metric/\\Accr.} & \multirow{2}{0cm}{eDF} & \multicolumn{4}{c}{$R_{\rm high}$}\\
		 && 1  &  10  &  20 & 40 \\\hline 		
		 	\multirow{3}{0cm}{Kerr\\SANE}		 & thermal          & $0.190\pm0.024$ & $6.10\pm0.80$  & $14.4\pm1.8$  & $20.9\pm2.6$ \\
		        & $\varepsilon=0$     & $0.190\pm0.024$ & $6.1\pm0.8$    & $14.6\pm1.8$  & $21.3\pm2.7$ \\ 
		 		& $\varepsilon=0.015$ & $0.190\pm0.024$ & $6.1\pm0.8$    & $14.6\pm1.8$  & $21.2\pm2.6$ \\ \hline
					
		 \multirow{3}{0cm}{Dilaton\\SANE}               & thermal          & $0.211\pm0.028$ & $9.0\pm1.2$    & $26.0\pm4.0$  & $41.0\pm6.0$ \\ 
		     & $\varepsilon=0$     & $0.211\pm0.028$ & $9.1\pm1.2$    & $26.9\pm4.0$  & $43.3\pm6.0$ \\  
		  & $\varepsilon=0.015$ & $0.211\pm0.028$ & $9.1\pm1.2$    & $26.8\pm4.0$  & $43.1\pm6.0$ \\ \hline
		            
		   \multirow{3}{0cm}{Kerr\\MAD}               & thermal          & $0.124\pm0.019$ & $1.27\pm0.20$  & $1.68\pm0.27$ & $0.90\pm0.14$\\ 
		           & $\varepsilon=0$     & $0.124\pm0.020$ & $1.29\pm0.21$  & $1.71\pm0.27$ & $0.90\pm0.15$\\	
		 	         & $\varepsilon=0.015$ & $0.124\pm0.020$ & $1.29\pm0.20$  & $1.70\pm0.27$ & $0.90\pm0.14$\\ \hline
		
		   \multirow{3}{0cm}{Dilaton\\MAD}               & thermal          & $0.127\pm0.018$ & $0.82\pm0.12$  & $1.06\pm0.15$ & $1.30\pm0.18$\\ 
		        & $\varepsilon=0$     & $0.127\pm0.018$ & $0.85\pm0.12$  & $1.13\pm0.16$ & $1.40\pm0.20$\\  
		               & $\varepsilon=0.015$ & $0.127\pm0.018$ & $0.85\pm0.12$  & $1.12\pm0.16$ & $1.39\pm0.20$\\
		 \hline
	\end{tabular}
	\end{adjustbox}
	\label{tab:mdot_grrt}
\end{table}

\vspace{-20pt}
\section{Comparison to a Schwarzschild black hole}\label{sec:schwarzschild}
In order to disentangle influences of spacetime, accretion model and emission model on the image morphology, we compare both the Kerr and the dilaton simulation to that of a Schwarzschild spacetime. Since the Schwarzschild metric does not contain a free parameter, it is not possible to match it to the Kerr or dilaton spacetime at a characteristic radius.  In order for the equilibrium solution to have the same inner radius and reach to the MAD state in the late evolution, the initial torus size and angular momenta for Schwarzschild simulations are higher than in the Kerr case, $l_{\rm torus,\ Schwarzschild}=4.84$ and $l_{\rm torus,\ Schwarzschild}=6.84$, for SANE and MAD respectively. For details on the effects of spin and accretion model on GRMHD initial conditions, see \cite{Fromm2021b}.

Figure \ref{fig:GRMHD_Schwarzschild} shows the electron temperature at $R_{\rm high}=40$ for both SANE and MAD in Kerr, dilaton and Schwarzschild spacetimes. In SANE, the Schwarzschild system shows little polar outflow, but a more extended, moderate-temperature disk wind compared to the Kerr and dilaton systems. Moving to the MAD regime, the Schwarzschild system develops an un-collimated, hot temperature outflow.

Indicated by the black solid line in Fig. \ref{fig:SED_SgrA}, the Schwarzschild SED with $R_{\rm high}=40$ is comparable to the Kerr and dilaton SEDs for $R_{\rm high}=10$, while for MAD it lies in between $R_{\rm high}=1$ and $R_{\rm high}>1$ curves. In the MAD regime, the Schwarzschild spacetime fits the near-infrared flux measurements best. The near-infrared spectral index differs by about 0.5 for the thermal Schwarzschild SED in SANE and MAD, while coinciding very well when including non-thermal electrons in either accretion model (Fig. \ref{fig:alphaNIR_MAD}).

\begin{figure*}
	\centering
	\includegraphics[width=0.495\textwidth]{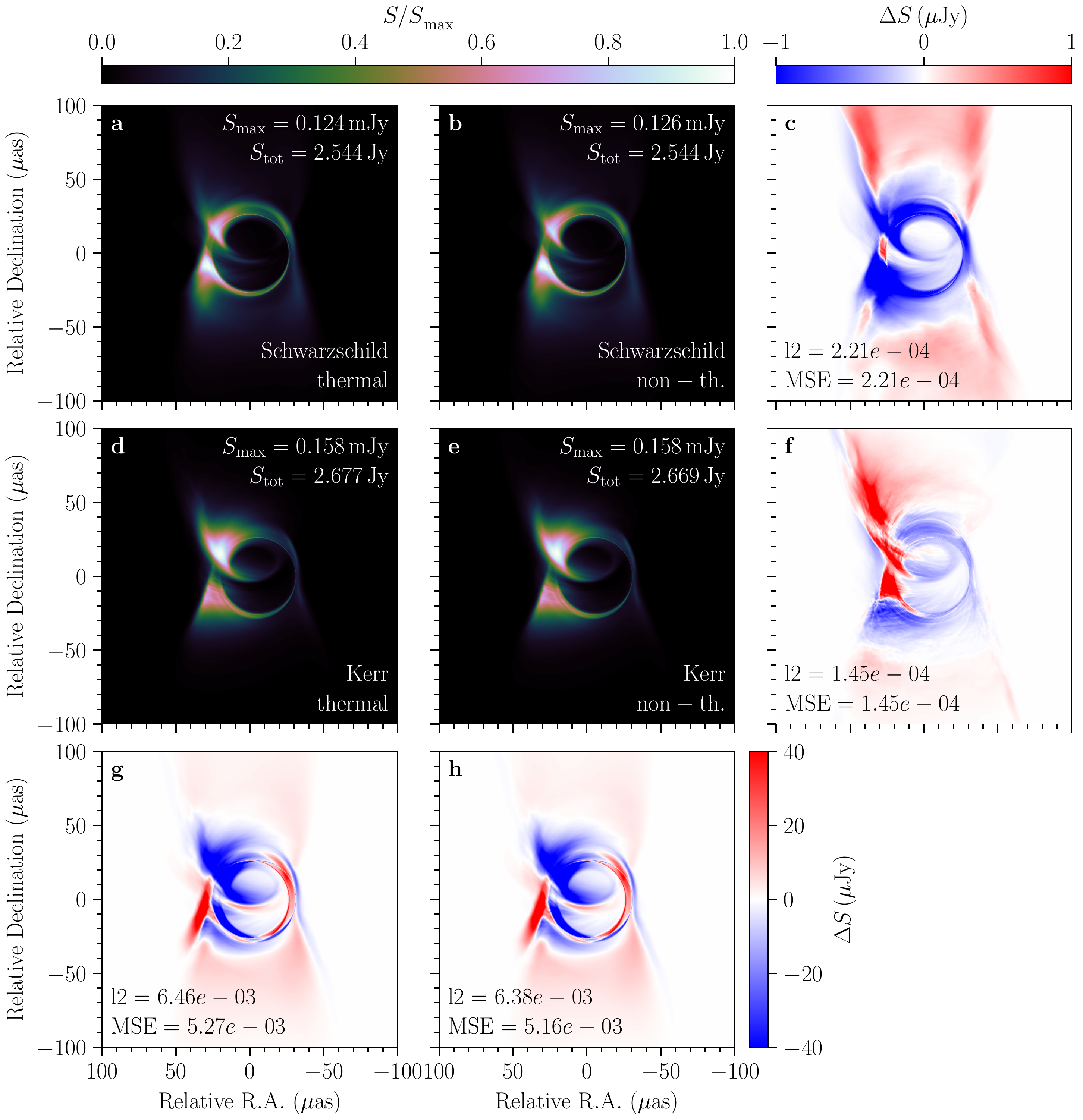}
	\includegraphics[width=0.495\textwidth]{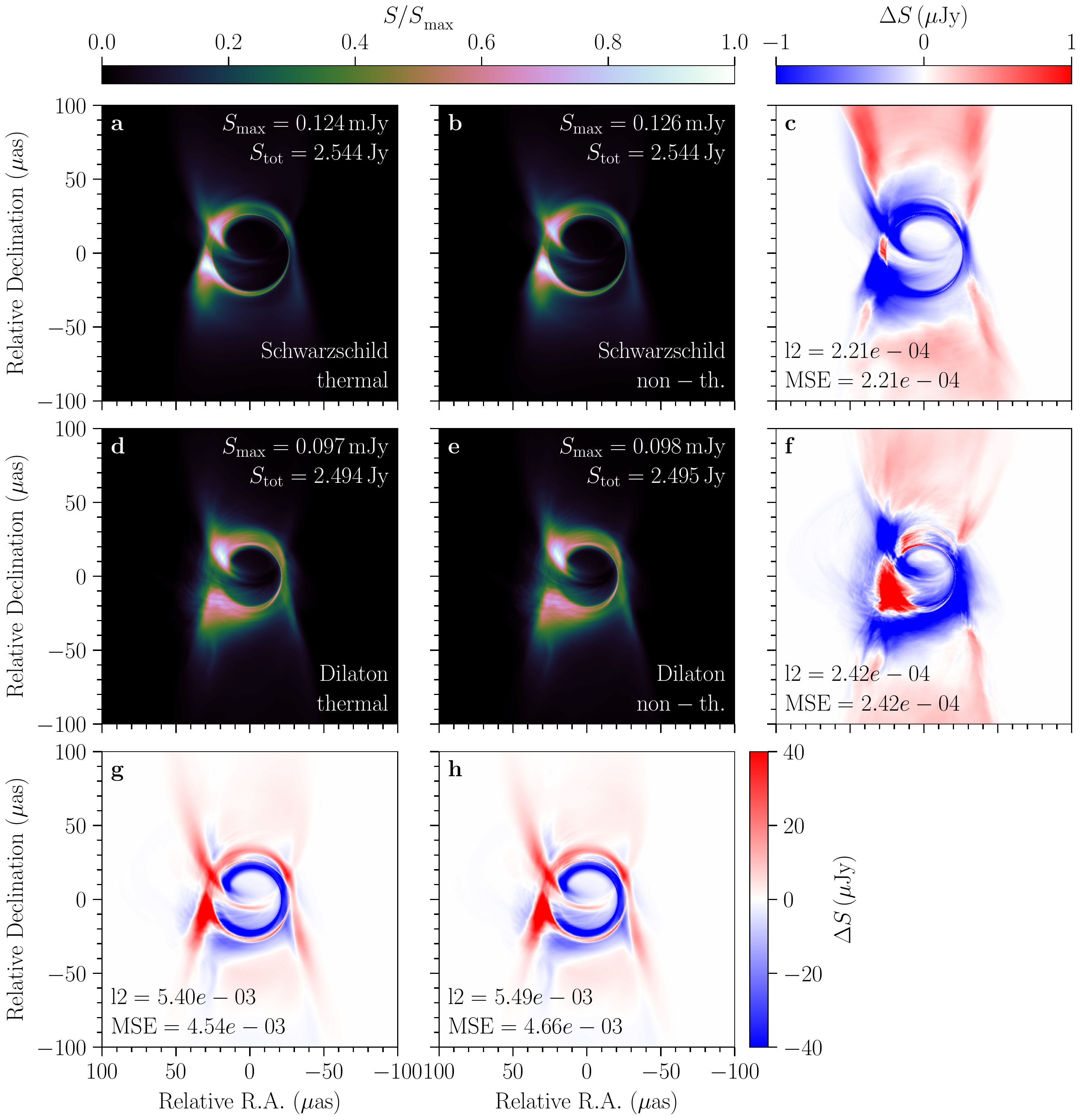}
	\caption{Left: comparison of Schwarzschild and Kerr spacetimes in thermal (left column) and non-thermal (middle column). Right: Comparison of the Schwarzschild to the dilaton spacetime. The images are averages of 100 snapshots over 1\,000\,M simulation time ($\sim$$6$\,h for Sgr\,A$^*$). A non-thermal electron distribution function has been applied in the jet sheath.} 
	\label{fig:GRRT_Schw_SANE}
\end{figure*}

\begin{figure*}
	\centering
	\includegraphics[width=0.495\textwidth]{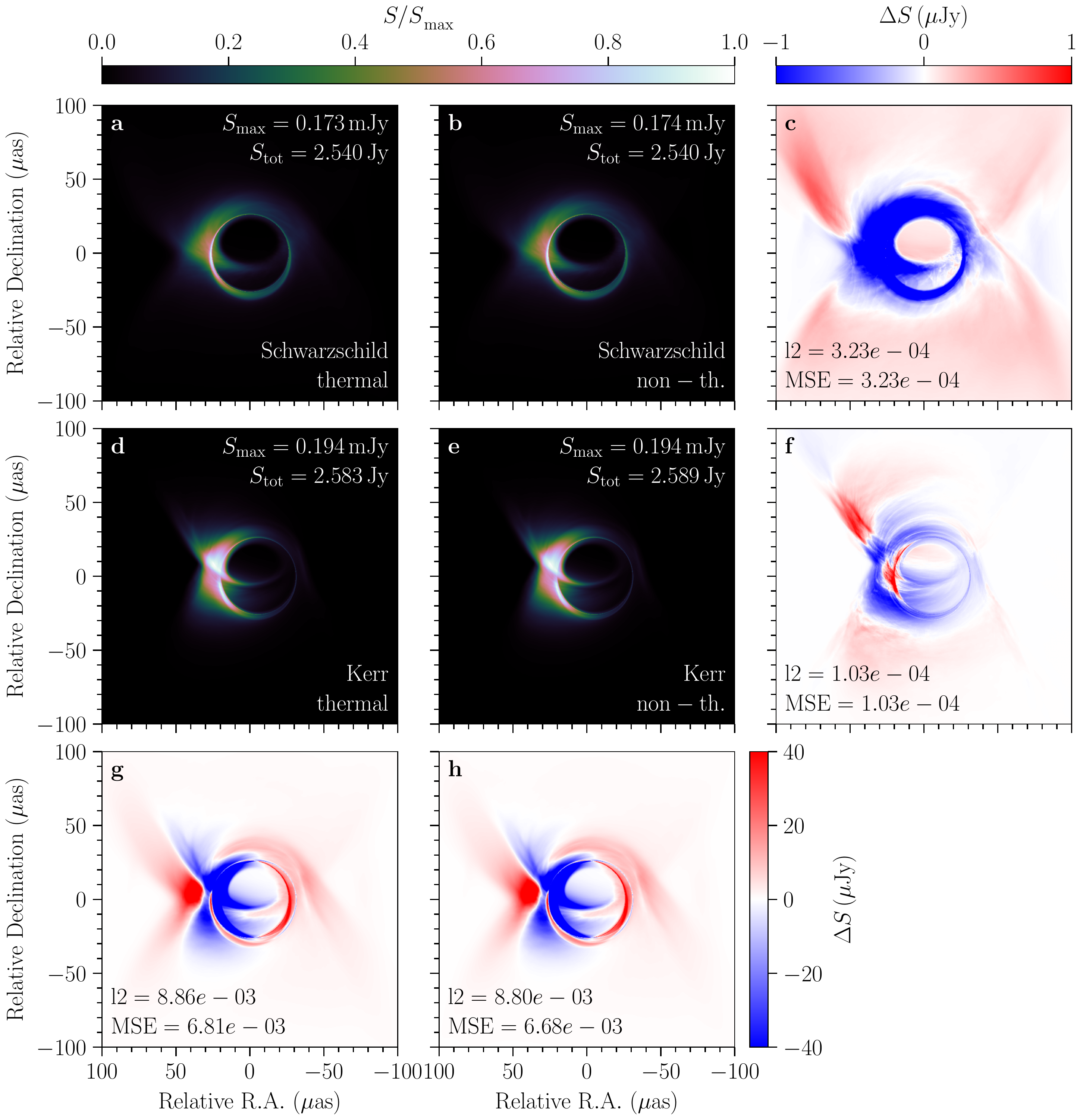}
	\includegraphics[width=0.495\textwidth]{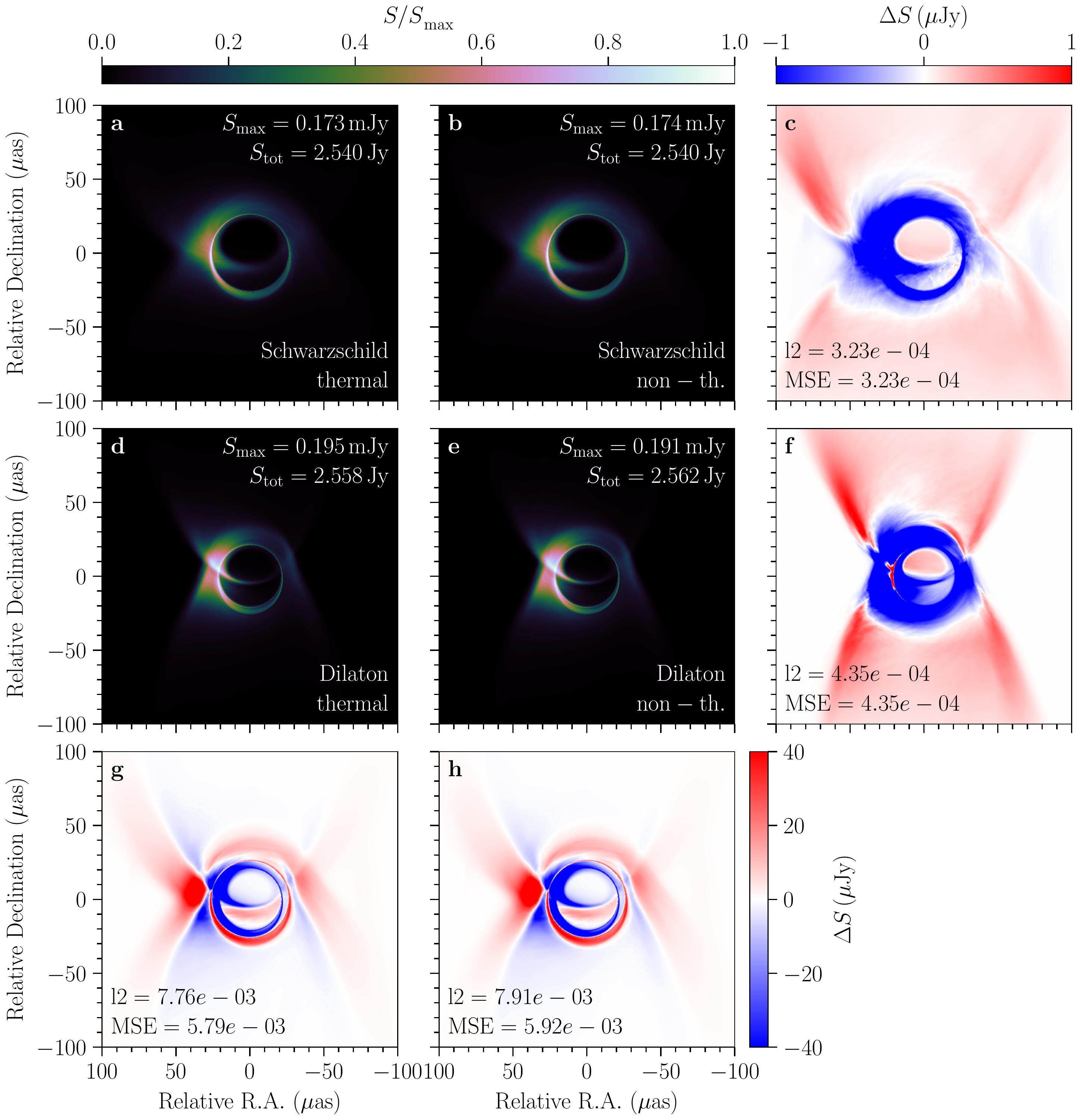}
	\caption{Same as Fig. \ref{fig:GRRT_Schw_SANE}, but for MAD.} 
	\label{fig:GRRT_Schw_MAD}
\end{figure*}

In the direct comparison of GRRT images, both the Schwarzschild and Kerr as well as the Schwarzschild and dilaton spacetimes differ by up to $\sim$$100$\,$\upmu$Jy (Figs. \ref{fig:GRRT_Schw_SANE} and \ref{fig:GRRT_Schw_MAD}). This is about a factor of two more than the differences between the Kerr and dilaton black holes, and can be attributed to the lack of matching between the black hole systems.

\newpage
\section{Multi-frequency observations of Sgr\,A$^\ast$}
In Table ~\ref{tab:obsSED} we provide the references to the observational data that is over-plotted on the broadband spectra in Fig. \ref{fig:SED_SgrA}. 

\begin{table*}[h!]
	\centering
	\def\arraystretch{1.0}
	\caption{Observed flux densities for Sgr\,A$^\ast$.}
	\label{tab:obsSED}
	\begin{adjustbox}{max width=0.88\textwidth}
		\begin{tabular}{cccll} 
			\hline
			\hline
			\bf{Frequency (GHz)} & \bf{Flux (Jy)} & \bf{Error (Jy)} & \bf{Instrument} &\bf{Reference}\\
			\hline
			\hline
			
			$136.3\times10^3$&0.0011&0.0003&GRAVITY&1\\\hline
			
			868.0&1.864&0.067&ALMA&2\\
			678.0&2.183&0.026&ALMA&2\\
			233.0&2.886&0.043&ALMA&2\\\hline
			
			$2.3\times10^3$&0.16&0.10&PACS (Herschel)&3\\
			$1.9\times10^3$&0.27&0.07&PACS (Herschel)&3\\	\hline	
			
			$136.3\times10^3$&0.002&0.005&VLT&4\\
			$66.6\times10^3$&0.0032&0.0034&IRAC (Spitzer)&4\\	\hline		
			
			$1.2\times10^3$&0.5&0.5&SPIRE (Herschel)&5\\\hline
			
			708.9&3.21&0.64&ALMA&6\\
			708.9&3.99&0.79&ALMA&6\\
			691.5&2.68&0.54&ALMA&6\\
			685.2&2.65&0.53&ALMA&6\\
			486.2&3.6&0.72&ALMA&7\\
			343.0&4.26&0.24&ALMA&6\\
			253.8&4.17&0.17&ALMA&6\\
			107.0&2.62&0.18&ALMA&6\\
			105.0&2.54&0.17&ALMA&6\\
			94.9&2.37&0.16&ALMA&6\\
			93.0&2.35&0.16&ALMA&6\\\hline
			
			216.8&3.677&0.762&SMA&8\\
			223.9&3.391&0.489&SMA&8\\
			238.2&3.31&0.424&SMA&8\\
			266.8&3.369&0.096&SMA&8\\
			274&3.526&0.697&SMA&8\\
			331.1&3.205&1.074&SMA&8\\
			338.3&3.436&0.863&SMA&8\\
			352.6&4.89&0.721&SMA&8\\
			218&3.667&0.65&ALMA&8\\
			220&3.661&0.652&ALMA&8\\
			231.9&3.676&0.664&ALMA&8\\
			233.8&3.704&0.68&ALMA&8\\
			341.6&3.602&0.866&ALMA&8\\
			343.6&3.609&0.87&ALMA&8\\
			351.7&3.595&0.884&ALMA&8\\
			353.6&3.553&0.86&ALMA&8\\\hline
			
			100.0&2.29&0.09&VLBA&9\\
			100.0&2.29&0.09&VLBA&9\\
			48.0&2&0.11&VLBA&9\\
			39.0&1.82&0.06&VLBA&9\\
			37.0&1.61&0.05&VLBA&9\\
			27.0&1.538&0.025&VLBA&9\\
			25.0&1.43&0.04&VLBA&9\\
			19.0&1.266&0.019&VLBA&9\\\hline
			
			$142.8\times10^3$&0.0022&0.0002&NIRC2 (Keck), NACO (VLT)&10\\
			$78.9\times10^3$&0.005&0.0006&VISIR (VLT)&10\\
			$62.5\times10^3$&0.0038&0.0013&VISIR (VLT)&10\\
			$34.9\times10^3$&0.084&0.04&VISIR (VLT)&10\\\hline
			
			$137.5\times10^3$&0.0011&n/a&NACO (VLT)&11\\	
			$25.2\times10^3$&0.057&0.03&VISIR (VLT)&12\\\hline
			
			344.6&6.7&1.5&SMA&13\\
			230.6&3.7&0.7&SMA&13\\
			99.9&1.3&0.6&NMA&13\\
			42.8&1.2&0.2&VLA&13\\
			23.1&1.15&0.17&VLA&13\\
			15.0&1.04&0.18&VLA&13\\
			
			\hline

		\end{tabular}

	\end{adjustbox}
\tablebib{(1) \citealt{Gravity2020c}; (2) \citealt{Bower2019}; (3) \citealt{vonFellenberg2018}; (4) \citealt{Witzel2018}; (5) \citealt{Stone2016}; (6) \citealt{Liu2016_1}; (7) \citealt{Liu2016_2}; (8) \citealt{Bower2015}; (9) \citealt{Brinkerink2015}; (10) \citealt{Schoedel2011}; (11) \citealt{Dodds-Eden2011}; (12) \citealt{Dodds-Eden2009}; (13) \citealt{Zhao2003}}
\end{table*}

\end{appendix}

\end{document}